\documentclass[aps,prx,twocolumn,a4paper,showpacs,superscriptaddress,longbibliography]{revtex4-1}

\usepackage{amsmath,amssymb,bm,graphicx}
\usepackage{xcolor,changes}
\usepackage[colorlinks=true,urlcolor=blue,linkcolor=blue,citecolor=blue,bookmarks=false]{hyperref}
\usepackage{ulem}

\begin{document}

\newcommand{\MG}[1]{\textcolor{red}{#1}}
\newcommand{\DG}[1]{\textcolor{green}{#1}}

\newcommand{\todo}[1]{\textbf{\textsc{\textcolor{red}{(TODO: #1)}}}}
\newcommand{\fcs}{Fe$_{1-x}$Co$_{x}$Si}
\newcommand{\mfs}{Mn$_{1-x}$Fe$_{x}$Si}
\newcommand{\mcs}{Mn$_{1-x}$Co$_{x}$Si}
\newcommand{\cso}{Cu$_{2}$OSeO$_{3}$}
\newcommand{\dmdh}{$\mathrm{d}M/\mathrm{d}H$}

\definechangesauthor[name=CP, color=red]{1}

%%%%%%%%%%%%%%%%%%%%%%%%%%%%%%%%%%%%%%%%%%%%%%%%%

\title{Weak crystallization of fluctuating skyrmion textures in MnSi}

\author{J. Kindervater}
\affiliation{Lehrstuhl f\"ur Topologie korrelierter Systeme, Physik-Department, Technische Universit\"at M\"unchen, D-85748 Garching, Germany}

\author{I. Stasinopoulos}
\affiliation{Lehrstuhl f\"ur Physik funktionaler Schichtsysteme, Physik-Department, Technische Universit\"at M\"unchen, D-85748 Garching, Germany}

\author{A. Bauer}
\email[e-mail:]{andreas.bauer@ph.tum.de}
\affiliation{Lehrstuhl f\"ur Topologie korrelierter Systeme, Physik-Department, Technische Universit\"at M\"unchen, D-85748 Garching, Germany}

\author{F. X. Haslbeck}
\affiliation{Lehrstuhl f\"ur Topologie korrelierter Systeme, Physik-Department, Technische Universit\"at M\"unchen, D-85748 Garching, Germany}

\author{F. Rucker}
\affiliation{Lehrstuhl f\"ur Topologie korrelierter Systeme, Physik-Department, Technische Universit\"at M\"unchen, D-85748 Garching, Germany}

\author{A. Chacon}
\affiliation{Lehrstuhl f\"ur Topologie korrelierter Systeme, Physik-Department, Technische Universit\"at M\"unchen, D-85748 Garching, Germany}

\author{S. M\"uhlbauer}
\affiliation{Heinz Maier-Leibnitz Zentrum (MLZ), Technische Universit\"at M\"unchen, D-85748 Garching, Germany}

\author{C. Franz}
\affiliation{Heinz Maier-Leibnitz Zentrum (MLZ), Technische Universit\"at M\"unchen, D-85748 Garching, Germany}

\author{M. Garst}
\affiliation{Institut f\"ur Theoretische Physik, Technische Universit\"at Dresden, 01062 Dresden, Germany}
\affiliation{Institut f\"ur Theoretische Festk\"orperphysik, Karlsruhe Institute of Technology, 76131 Karlsruhe, Germany}

\author{D. Grundler}
\affiliation{Lehrstuhl f\"ur Physik funktionaler Schichtsysteme, Physik-Department, Technische Universit\"at M\"unchen, D-85748 Garching, Germany}
\affiliation{Institute of Materials and Laboratory of Nanoscale Magnetic Materials and Magnonics, \'Ecole Polytechnique F\'ed\'erale de Lausanne, CH-1015 Lausanne, Switzerland}

\author{C. Pfleiderer}
\email[e-mail:]{christian.pfleiderer@tum.de}
\affiliation{Lehrstuhl f\"ur Topologie korrelierter Systeme, Physik-Department, Technische Universit\"at M\"unchen, D-85748 Garching, Germany}

\date{\today}

\begin{abstract}
We report an experimental study of the emergence of non-trivial topological winding and long-range order across the paramagnetic to skyrmion lattice transition in the transition metal helimagnet MnSi. Combining measurements of the susceptibility with small angle neutron scattering, neutron resonance spin echo spectroscopy and all-electrical microwave spectroscopy, we find evidence of skyrmion textures in the paramagnetic state exceeding $10^3\,{\rm \AA}$ with lifetimes above several 10$^{-9}$\,s. Our experimental findings establish that the paramagnetic to skyrmion lattice transition in MnSi is well-described by the Landau soft-mode mechanism of weak crystallization, originally proposed in the context of the liquid to crystal transition. As a key aspect of this theoretical model, the modulation-vectors of periodic small amplitude components of the magnetization form triangles that add to zero. In excellent agreement with our experimental findings, these triangles of the modulation-vectors entail the presence of the non-trivial topological winding of skyrmions already in the paramagnetic state of MnSi when approaching the skyrmion lattice transition. 
\end{abstract}

\maketitle

%%%%%%%%%%%%%%%%%%%%%%%%%%%%%%%%%%%%%%%%%%%%%%%%%

\section{Motivation}

A pre-requisite for the definition of topological magnetic textures is the presence of a 
continuous magnetization field with a finite amplitude in space and time. An example \textit{par excellence} of such textures are magnetic skyrmions, representing topologically nontrivial whirls of this magnetization field~\cite{2013_Nagaosa_NatNanotechnol}. The notions of topological winding and topological stability of such skyrmions are only meaningful when the magnetization is sufficiently smooth on local scales. This condition may be readily satisfied in systems exhibiting long-range magnetic order for temperatures far below the transition temperature $T_c$. In contrast, changes of the topological properties require that the magnetization is capable of vanishing on short length and time scales. The associated microscopic mechanisms underlying the transition of skyrmions into different types of conventional long-range magnetic order have been explored in a large number of theoretical and experimental studies~\cite{2013_Milde_Science, 2017_Wild_SciAdv, 2015_Rajeswari_ProcNatlAcadSciUSA, 2018_Berruto_PhysRevLett, 2013_Romming_Science, 2017_Hsu_NatNanotechnol, 2015_Jiang_Science}.

A major unresolved question concerns, in contrast, the formation of skyrmion lattice order when starting from a state that is essentially paramagnetic and dominated by an abundance of fluctuations such that the local magnetization, on a coarse-grained level, practically vanishes on average \cite{1989_Bogdanov_SovPhysJETP, 1994_Bogdanov_JMagnMagnMater, 1999_Bogdanov_JMagnMagnMater, 2006_Rossler_Nature, 2011_Wilhelm_PhysRevLett, 2013_Finazzi_PhysRevLett, 2016_Bauer_Book}.  This alludes to the question whether topologically non-trivial characteristics exist already in a paramagnetic state, and how they may be accounted for in the framework of the present-day classification of phase transitions~\cite{1971_Stanley_Book, 1983_Wilson_RevModPhys, 1992_Binney_Book}. It connects also with the relevance of non-trivial topological properties in the search for novel electronic properties of solids, e.g., at quantum phase transitions~\cite{2007_Lohneysen_RevModPhys, 2008_Sachdev_NatPhys}.

Magnetic skyrmions are ideally suited to address this question. However, they are typically portrayed from either one of two seemingly contrary points of view. On the one hand, skyrmions may be described as solitonic objects in real space with a well-defined particle-like appearance \cite{1989_Bogdanov_SovPhysJETP, 1994_Bogdanov_JMagnMagnMater, 1999_Bogdanov_JMagnMagnMater, 2006_Rossler_Nature, 2015_Romming_PhysRevLett, 2010_Yu_Nature}, which in the extreme limit may be reminiscent of hard spheres. On the other hand, skyrmions may be described in terms of smooth, wave-like textures of the magnetization field~\cite{2009_Muhlbauer_Science, 2013_Buhrandt_PhysRevB}, where long-range skyrmion lattices form multi-dimensionally modulated states, also known as multi-$\vec{Q}$ magnetic order. This suggests rather different scenarios for the emergence of long-range skyrmion lattice order.

With the emergence of the skyrmion lattice the translation symmetry is broken. This symmetry breaking may be linked more strongly or more weakly to the appearance of skyrmionic entities near the transition. Consider the extreme limit of a strong first-order transition where the magnetization and the non-trivial topological winding would arise simultaneously out of a pristine mean-field like paramagnetic state. In this situations, one would not expect any skyrmion-like precursor phenomena in the paramagnetic phase. In the other extreme limit, particle-like skyrmionic solitons might be preformed in the paramagnet, and the transition to long-range skyrmion lattice order may be expected to share similarities with the crystallization of atoms or molecules in a liquid. Here, the non-trivial topological winding of the skyrmions as a key ingredient of the ordered phase, would already be present in the paramagnetic state. 

An intermediate situation might arise in case of transitions that are only very weakly first order. In the spirit of the theory of weak crystallization originally put forward  by Landau in 1937~\cite{1937_Landau_ZhEkspTeorFiza, 1987_Brazovskii_SovPhysJETP}, the free energy of the system may then be expanded in those Fourier components of the magnetization which belong to the reciprocal lattice of the emergent skyrmion crystal. In the presence of a finite field, cubic terms are allowed that lower the energy efficiently when there exist modulation vectors within a plane that add to zero, i.e., when they form triangles, which eventually leads to the stabilization of hexagonal skyrmion lattice order. In the paramagnetic state close to the transition, the fluctuations of Fourier components with triangular wavevector configurations might then give rise to patches with skyrmionic character which have not yet developed static long-range order. 

In recent years a large number of experimental and theoretical studies have shown that magnetic skyrmions are ubiquitous in bulk compounds, thin films, heterostructures, and nano-scaled systems, where both particle-like and wave-like descriptions have been used~\cite{1989_Bogdanov_SovPhysJETP, 1994_Bogdanov_JMagnMagnMater, 1999_Bogdanov_JMagnMagnMater, 2009_Muhlbauer_Science, 2010_Yu_Nature, 2011_Yu_NatMater, 2012_Seki_Science, 2015_Kezsmarki_NatMater, 2015_Tokunaga_NatCommun, 2017_Nayak_Nature, 2011_Heinze_NatPhys, 2013_Romming_Science, 2016_Wiesendanger_NatRevMater, 2013_Finazzi_PhysRevLett, 2015_Buttner_NatPhys, 2016_Woo_NatMater, 2015_Jiang_Science, 2017_Jiang_NatPhys, 2013_Sampaio_NatNanotechnol, 2016_Moreau-Luchaire_NatNanotechnol, 2018_Maccariello_NatNanotechnol, 2017_Fert_NatRevMater}. Despite this enormous progress the question of the emergence of skyrmions from an environment that is strongly fluctuating and essentially paramagnetic has not been resolved. Combining high-precision AC susceptibility measurements with small-angle neutron scattering, neutron spin echo spectroscopy, and ferromagnetic resonance studies we resolve this question for the skyrmion lattice order in MnSi. The order emerges from a disordered state in terms of a weak crystallization of fluctuating skyrmion textures with topologically non-trivial character.

%%%%%%%%%%%%%%%%%%%%%%%%%%%%%%%%%%%%%%%%%%%%%%%%%
\section{State of the Art and Main Results}

Historically first, and so far most extensively studied are skyrmions in the class of cubic chiral magnets, where the interactions feature a hierarchy of scales that comprises in decreasing strength ferromagnetic exchange interactions, Dzyaloshinsky-Moriya (DM) spin-orbit coupling and magnetocrystalline anisotropies~\cite{1984_Landau_Book}. As a consequence of the strongest scale, cubic chiral magnets exhibit paramagnetic behavior at high temperatures and low fields, and ferromagnetic behavior at high fields and low temperatures. Poised at the border between the paramagnetic and ferromagnetic regimes in the limit of low temperatures and low magnetic fields is a complex magnetic phase diagram featuring long-wavelength chiral modulations of the magnetization that reflect the DM interactions and magneto-crystalline anisotropies \cite{1976_Kasuka_SSC, 1977_Ishikawa_PhysicaB, 2012_Bauer_PhysRevB, 2016_Bauer_Book}. 

At zero magnetic field the modulated magnetic state is a long-wavelength helical modulation, where an applied magnetic field generates a reorientation of the direction of the helimagnetic modulation at a field $H_{c1}$ such that the so-called conical state stabilizes. This is followed by a transition to the field-polarized (ferromagnetic) state when increasing the field above $H_{c2}$. The skyrmion lattice phase is, finally, embedded in the conical state under a small applied magnetic field in the vicinity of the paramagnetic to helimagnetic transition. 

A series of studies have established that the magnetic phase diagram may be fully accounted for in terms of a conventional Ginzburg-Landau theory with fluctuation corrections. In particular, it has long been recognized that the long-range skyrmion lattice in cubic chiral magnets may be approximately described by a triple-$\vec{Q}$ structure in a plane perpendicular to an applied magnetic field, where the wave-vectors enclose rigid angles of 120$^{\circ}$~\cite{2009_Muhlbauer_Science, 2013_Buhrandt_PhysRevB}. Analysis of the spin orientation established a non-zero topological winding number per magnetic unit cell characteristic of skyrmions. For the thermodynamic stability of the skyrmion lattice in the tiny phase pocket near $T_c$ thermal fluctuations are essential~\cite{2009_Muhlbauer_Science, 2013_Buhrandt_PhysRevB}.

In turn, when starting in the skyrmion lattice phase a decrease of temperature or changes of the applied magnetic field cause a transition of the skyrmion lattice to the conical state. Moreover, under thermal quenches in an applied magnetic field the skyrmion lattice survives as a metastable state down to very low temperatures, where a direct transition into the helical or field-polarized phase may be triggered in field-sweeps \cite{2013_Milde_Science, 2017_Wild_SciAdv, 2016_Oike_NatPhys, 2017_Nakajima_SciAdv, 2018_Nakajima_PhysRevB,2018_Chai_ArXiv}. 

In contrast, the skyrmion lattice undergoes a transition to a paramagnetic state when increasing the temperature. Even though there exists a small field-induced uniform magnetization above $T_c$ we denote this state as paramagnetic, because it features an abundance of fluctuations and  components of the magnetization at and around the ordering vector $\vec{Q}$ that are purely dynamic. Conceptually, three contributions to the spectrum of fluctuations may be distinguished: (i) components dominated by the magnetocrystalline anisotropies that are reminiscent of the helical order, (ii) components dominated by the Zeeman energy akin to the conical order parallel to the applied field, and (iii) components in the plane perpendicular to the applied field featuring a multi-$Q$ character akin to the skyrmion lattice. Delineation of different contributions at the paramagnetic to skyrmion lattice transition represents an important question.

On a more general note, the paramagnetic to skyrmion lattice transition in MnSi is of interest in the search for novel electronic properties of solids, where a non-Fermi liquid resistivity and concomitant topological Hall signal has been observed in MnSi under high pressure as well as in {\mfs} under substitutional Fe doping \cite{2001_Pfleiderer_Nature, 2009_Lee_PhysRevLett, 2013_Chapman_PhysRevB, 2013_Ritz_PhysRevB, 2013_Ritz_Nature, 2014_Franz_PhysRevLett, 2019_Uemura_QuanMat}. These studies raise the question for microscopic evidence of slowly varying spin textures and topological winding in the paramagnetic state.

An important point of reference for the skyrmion lattice to paramagnetic transition at finite applied magnetic field and ambient pressure is the behavior in zero magnetic field~\cite{2005_Grigoriev_PhysRevB, 2013_Janoschek_PhysRevB, 2007_Stishov_PhysRevB}. A peak-hump temperature dependence of the specific heat in MnSi \cite{2001_Pfleiderer_JMMM} inspired theoretical work taking into account a non-analytic gradient term beyond conventional Ginzburg-Landau theory, that suggested the putative existence of a spontaneous (zero field) skyrmion phase between the paramagnetic and helimagnetic state \cite{2006_Rossler_Nature}. While initial small angle neutron scattering and thermal expansion measurements did not favor such a skyrmion phase \cite{2005_Grigoriev_PhysRevB, 2007_Stishov_PhysRevB}, polarized neutron scattering data were interpreted in support of this prediction \cite{2009_Pappas_PhysRevLett, 2011_Pappas_PhysRevB,2013_Shan_pss}.

Comprehensive SANS, neutron polarimetry, specific heat and susceptibility measurements eventually ruled out the existence of a skyrmion liquid at zero field in MnSi \cite{2013_Janoschek_PhysRevB,2014_Kindervater_PhysRevB}. Namely, when approaching $T_c$ the fluctuations in the paramagnetic state develop a helimagnetic character. Close to $T_c$, the weight of the magnetic structure factor accumulates on a sphere in wavevector space with a radius $Q$ that is determined by the competition between exchange and DM interaction. Such a weight distribution implies strong renormalization effects. All of the experimental findings could be described in the framework of conventional Landau theory without need for the non-analytic term. The account is consistent with the scenario of a fluctuation-induced first-order transition proposed by Brazovskii, where the paramagnetic-to-helimagnetic transition is weakly first order with a peak-hump structure in the specific heat.
%and related properties.

Application of a magnetic field quenches the fluctuations resulting in a tricritical point at ${\sim}0.4$~T~\cite{2013_Bauer_PhysRevLett, 2014_Nii_PhysRevLett}. This raises the question for the evolution of the transition to long range magnetic order as a function of temperature under different magnetic fields, namely variations of helical, conical and skyrmionic components of the fluctuations as mentioned above. To resolve this issue requires, (i) to collect comprehensive information of the distribution and character of the spectrum of spin fluctuations throughout reciprocal space, (ii) to delineate single-$Q$ from multi-$Q$ fluctuations, and (iii) to connect the fluctuations quantitatively with bulk properties, notably magnetization, susceptibility, and specific heat. The associated experimental and theoretical work is clearly prohibitive in terms of the necessary resources and time, and thus well-beyond present day capabilities.

In contrast, a well-defined key question that may be resolved concerns the character of the paramagnetic state and the emergence of the non-trivial topological winding in the skymrion lattice phase when approaching the skyrmion lattice transition under decreasing temperature. Based on the AC susceptibility and limited specific heat data at finite field, it was argued that skyrmionic precursor phenomena exist in the paramagnetic state at the border of the skyrmion lattice phase in FeGe \cite{2011_Wilhelm_PhysRevLett, 2013_Cevey_pssb}. However, this interpretation lacked the necessary microscopic information on the spatial, temporal and topological character of the correlations. Moreover, the claim of skyrmionic precursor phenomena in the paramagnetic state assumed a spontaneous skyrmion liquid in zero magnetic field. Thus, in order to clarify the nature of the paramagnetic to skyrmion lattice transition a combination of bulk properties and microscopic data is required, as well as a critical assessment of the consistency with conventional Ginzburg-Landau theory.

In this paper we report an experimental investigation of the paramagnetic to skyrmion lattice transition in MnSi, the most extensively studied representative of the class of cubic chiral magnets. In our study we combine the information inferred from four experimental methods as described in section \ref{methods}. This permits to show, that the paramagnetic to skyrmion lattice transition is dominated by skyrmionic fluctuations, whereas conical and helical fluctuations are subleading. In turn, an assessment of the full magnetic field and temperature dependence of the spectrum of fluctuations by means of neutron scattering, which is not feasible to date, is not required.

Our arguments and presentation are organized as follows. High-precision measurements of the longitudinal and transverse ac susceptibility, reported in section \ref{ACS}, reveal that remnants of key signatures attributed to skyrmion lattice order extend into the paramagnetic regime up to few K above the critical temperature. As presented in section \ref{SANS}, in this paramagnetic regime an abundance of fluctuations gives rise to small angle neutron scattering intensity on the surface of a sphere with a faint six-fold intensity pattern reminiscent of the skyrmion lattice order, where the integrated weight is dominant. This establishes that further contributions with a conical character are subleading. The associated correlation lengths of this skyrmionic intensity pattern are resolution-limited and exceed several thousand \AA. Thus, the six-fold pattern is quantitatively consistent with the strength of the relevant magneto-crystalline anisotropies that orient the pattern, which are sixth order in spin-orbit coupling [cf. section \ref{scales}]. 

Using neutron spin echo spectroscopy, we establish next that the paramagnetic regime is fully dynamic, where typical life-times of the skyrmionic correlations exceed several $10^{-9}\,$s, as reported in section \ref{MIEZE}. Finally, using microwave spectroscopy, reported in section \ref{FMR}, we find within this small temperature range the characteristics of counter-clockwise collective excitations of skyrmions around 10\,GHz, i.e., on time-scales that are short as compared with the lifetimes of the correlations. This underscores that skyrmionic correlations are dominant in comparison to conical contributions as seen in the other properties, in particular SANS.

As summarized in section \ref{summary}, the main experimental result of our paper concerns the observation of large-scale, long-lived fluctuations with skyrmionic character in the paramagnetic state of MnSi when approaching the skyrmion lattice transition. Evidence of the non-trivial topological character is provided by magnetic resonances that are reminiscent of typical excitations of the skyrmion lattice. Closer inspection of the length and time-scales inherent to our data presented in section \ref{scales}, such as the faint sixfold intensity pattern in the paramagnetic regime, are in excellent quantitative agreement with the well-known hierarchy of scales at the heart of all of the properties of MnSi and related compounds reported in the literature.

The presence of such large and slowly fluctuating skyrmion textures would naturally explain the topological Hall signal in MnSi seen under high pressures and substitutional Fe-doping as discussed in section \ref{emergent}. In Fourier space, these magnetic fluctuations correspond to the leading Fourier components of the emerging hexagonal skyrmion lattice with wavevectors that form triangles [cf. section \ref{weak-cryst}]. This identifies the paramagnetic to skyrmion lattice transition as a form of weak crystallization in the spirit of Landau's seminal proposal. Perhaps most remarkably, the large fluctuating skyrmion textures establish the presence of non-trivial topological winding in the disordered state.

%%%%%%%%%%%%%%%%%%%%%%%%%%%%%%%%%%%%%%%%%%%%%%%%%
%\newpage
\section{Experimental Methods}
\label{methods}

For our study high-quality single-crystal samples of MnSi were grown by means of optical float-zoning under ultra-high vacuum compatible conditions~\cite{2011_Neubauer_RevSciInstrum, 2016_Bauer_RevSciInstruma}. The AC susceptibility and specific heat was measured on a cubic sample with an edge length of 2~mm and surfaces perpendicular to $\langle100\rangle$ and $\langle110\rangle$. For the small-angle neutron scattering a spherical single crystal with a diameter of 5.75~mm was used, where a crystallographic $\langle110\rangle$ direction was perpendicular to the directions of the neutron beam and the magnetic field. The neutron resonance spin echo spectroscopy was performed on a cylindrical MnSi sample with diameter of 10~mm and a length of 30~mm, where the $\langle110\rangle$ direction was parallel to the symmetry axis of the cylinder. For the microwave spectroscopy a disc was used with a diameter of 6~mm and a height of 1~mm, corresponding to demagnetization factors $N_{x} = N_{y} = 0.175$ and $N_{z} = 0.651$. 

The longitudinal AC susceptibility was measured with a Quantum Design physical properties measurement system. For the measurements of the transverse AC susceptibility, a bespoke susceptometer was used~\cite{2017_Rucker_PhD, 2019_Rucker_RevSciInstrum} as operated in a superconducting magnet system with a variable temperature insert. Both, the longitudinal and transverse susceptibility were measured at an excitation frequency of 120~Hz and an excitation amplitude of 0.5~mT. Complimentary magnetization and specific heat data, recorded in order to confirm the sample quality with the literature and previous studies, were also determined using the Quantum Design physical properties measurement system.

The SANS measurements were performed at SANS-1~\cite{2015_Heinemann_JLarge-ScaleResFacil} at the Heinz Maier-Leibnitz Zentrum~(MLZ) in Garching. Data were recorded for a neutron wavelength of $\lambda = 5.5~{\rm \AA}$ with a wavelength spread $\Delta\lambda/\lambda\sim0.1$. The neutron beam was collimated over a length of 12~m with a beam diameter of 50~mm. 350~mm in front of the sample a pinhole aperture was placed with a diameter of 4~mm. The sample detector distance was 10~m. 

For an estimate of the resolution of the SANS measurements we assume a Gaussian distribution of the wavelength spread and beam divergence, which in reality are triangular and trapezoidal, respectively.  No collimating effects of the sample may be expected as the sample is larger than the pinhole aperture. Taking additionally the size of the pixels of the detector into account, the combined calculated resolution of our SANS set-up is given by a Gaussian with a full-width-half-maximum $\Delta_q\,= 0.0068\,{\rm \AA}^{-1}$. Due to these approximations the calculated resolution represents a conservative upper limit, whereas the measured full-width-half-maximum of the resolution is, in fact, smaller and given by $w_{\rm G}\,= 0.0051\,{\rm \AA}^{-1}$.

In the SANS studies magnetic fields were generated with a 5~T superconducting magnet system and the sample cooled with a pulse tube cooler. At the time the SANS data was recorded, the SANS-1 beam-line was not equipped with a goniometer. Therefore, all data were recorded using one-axis rocking scans with respect to the vertical direction. However, a large number of related studies using full two-axis rocking scans confirmed the full symmetry of the scattering pattern. Based on the very systematic temperature and magnetic field protocol only data recorded using one-axis rocking scans are shown in the following. We note that the logarithmic color-scales used in Fig.~\ref{figure2} are chosen such as to minimize the two-fold appearance of the diffraction pattern. 

Longitudinal neutron resonance spin echo spectroscopy was performed at the beam-line RESEDA~\cite{2015_Franz_JLarge-ScaleResFacil, 2019_Franz_NuclInstrumMethodsPhysResA, 2019_Franz_JPhysSocJpn} at the MLZ. As our study pursued measurements under applied magnetic fields the instrument was operated in the so-called modulation-of-intensity-with-zero effort (MIEZE) mode. In this mode the signal contrast corresponds to the intermediate scattering function $S(q,\tau)$. For a pedagogical account of the MIEZE set-up used in our study we refer to Refs.~\cite{2011_Georgii_ApplPhysLett, 2016_Krautloher_RevSciInstrum}. A neutron wavelength of $\lambda = 6$~\AA\ was used and data were recorded covering a dynamic range from 70~ps to 1.9~ns. 

All-electrical microwave spectroscopy was performed by means of a coplanar waveguide~(CPW). The surface of the sample that was placed on the CPW was carefully polished. Earlier work focussing on the universal character of the FMR spectra in the ordered state of different materials systems have long established that different excitations modes may be clearly discerned as comprehensively illustrated, e.g., in Fig.\,5 in the supplement of Ref.\,\cite{2015_Schwarze_NatMater}. For the work reported in this paper the detection electronics and impedance matching were optimized further, where additional details and illustrations of the noise level using an identical set-up may be found in Refs.\,\cite{2017_Stasinopoulos_ApplPhysLett,2017_Stasinopoulos_SciRep}. The static magnetic field was oriented along the $z$ axis perpendicular to the CPW. For our set-up the strongest high-frequency component of the CPW induced the precessional motion of the spins along the $x$ axis. Data were recorded with a vector network analyzer, providing the relative amplitude of the scattering parameter $|\Delta S_{12}|$. For subtraction of the signal background, spectra were at first recorded at each given temperature for a magnetic field of 2~T not containing the magnetic resonance of interest. This was followed by a subtraction of data recorded at a high temperature of 35~K. 

Throughout our study great care was exercised to keep track of differences of characteristic field values due to differences of sample shape and the associated demagnetizing fields. Moreover, taking into account an excessive body of experimental data that were recorded in the context of other studies, the size of systematic differences of sample temperature between the different experimental apparatus was tracked and found to be very small. This allowed us to ascertain internal consistency of the data presented in this paper, as well as consistency between different quantities, notably the AC susceptibility, small-angle neutron scattering, neutron resonance spin echo spectroscopy, and microwave spectroscopy.

%%%%%%%%%%%%%%%%%%%%%%%%%%%%%%%%%%%%%%%%%%%%%%%%%

\section{Experimental Results}
\label{results}

The presentation of our experimental results is organized in four subsections: magnetic susceptibility in section \ref{ACS}, small angle neutron scattering in section \ref{SANS}, neutron resonance spin echo spectroscopy in section \ref{MIEZE}, and microwave spectroscopy in section \ref{FMR}.

\subsection{Magnetic Susceptibility}
\label{ACS}

\begin{figure*}
\includegraphics[width=1.0\linewidth]{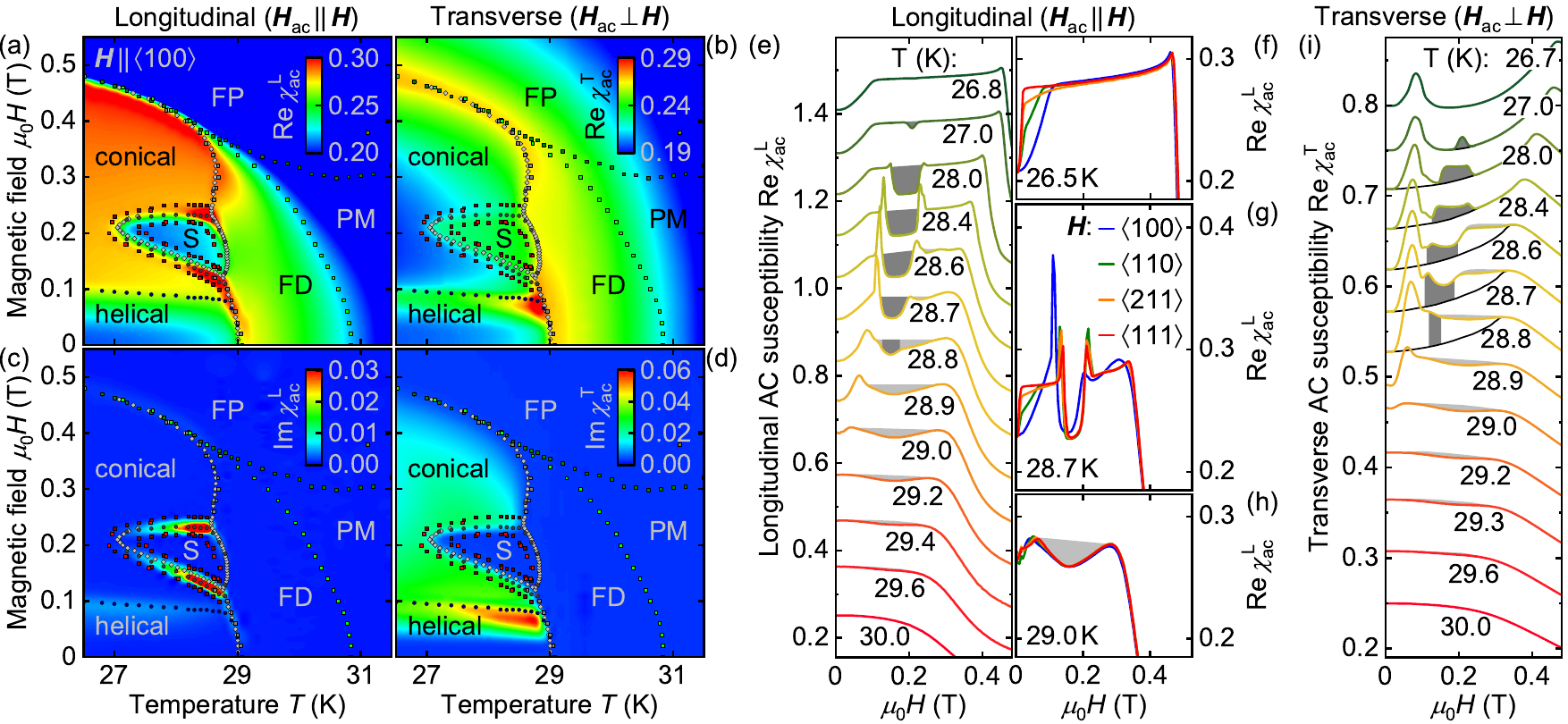}
\caption{Longitudinal and transverse ac susceptibility. \mbox{(a)--(d)}~Colormaps of the real and imaginary part of the longitudinal and transverse susceptibility, $\chi_{\mathrm{ac}}^{\mathrm{L}}$ and $\chi_{\mathrm{ac}}^{\mathrm{T}}$, for field along $\langle100\rangle$ after zero-field cooling. Data points are inferred from temperature (light colors) and field sweeps (dark colors) of the differential susceptibility $\mathrm{d}M/\mathrm{d}H$ (circles), $\chi_{\mathrm{ac}}^{\mathrm{L}}$ (squares), and specific heat measurements (diamonds); see text for details. The following six regimes may be distinguished: helical, conical, skyrmion lattice~(S), fluctuation-disordered~(FD), paramagnetic~(PM), field-polarized~(FP). (e)~$\mathrm{Re}\,\chi_{\mathrm{ac}}^{\mathrm{L}}$ as a function of field for different temperatures. Dark and light shadings indicate contributions that arise from the skyrmion lattice state and skyrmionic fluctuations, respectively. Data have been offset for clarity. \mbox{(f)--(h)}~$\mathrm{Re}\,\chi_{\mathrm{ac}}$ for different field directions and temperatures well below, around, and just above $T_{c}$. Isotropic behavior is observed in the regime without static order. (i)~$\mathrm{Re}\,\chi_{\mathrm{ac}}^{\mathrm{T}}$ as a function of field for different temperatures.}
\label{figure1}
\end{figure*}

It is helpful to begin with a brief reminder of key characteristics of the magnetic phase diagram of MnSi [see Figs\,\ref{figure1}\,(a) -- (d) and \ref{figure7}\,(a)].  As emphasized in the introduction two major regimes may be distinguished, notably large temperatures and small fields vis a vis low temperatures and large fields. While the former is essentially paramagnetic (PM) with a pronounced Curie-Weiss susceptibility of large fluctuating moments $\mu_{\rm CW}\approx2.2\mu_{\rm B}$, the latter is essentially ferromagnetic (field-polarized: FP). In MnSi the magnetic moment in the FP regime is strongly reduced, $\mu_{\rm s}\approx0.4\mu_{\rm B}$ as compared to the Curie-Weiss moment, and highly unsaturated under large magnetic fields \cite{1975_Bloch_PLA}. Both aspects represent a key characteristic of itinerant-electron magnetism~\cite{1985_Lonzarich_JPhysC, 1985_Moriya_Book}.

At zero magnetic field helimagnetic order with an ordering wave vector $\vec Q$ with $|\vec Q|=0.036\,{\rm \AA^{-1}}$ emerges below a transition temperature $T_c=29.0\,{\rm K}$. Detailed SANS, susceptibility, specific heat, ultra-sound attenuation, and thermal expansion data of the transition consistently establish a fluctuation-induced first order transition~\cite{2007_Stishov_PhysRevB, 2013_Janoschek_PhysRevB, 2013_Bauer_PhysRevLett, 2014_Nii_PhysRevLett, 2017_Pappas_PhysRevLett, 2019_Janoschek_PhysRevLett}, where the paramagnetic state when approaching $T_c$ develops an abundance of chiral fluctuations in a small temperature interval, denoted fluctuation-disordered (FD). 

At low temperatures a magnetic field exceeding a characteristic field $H_{c1}$ causes a reorientation of $\vec{Q}$ to become parallel to the applied field~\cite{2006_Grigoriev_PhysRevB, 2017_Bauer_PhysRevB}. Depending on crystallographic direction the reorientation at $H_{c1}$ is either a first or second order transition or a cross-over~\cite{2017_Bauer_PhysRevB}. When further increasing the magnetic field the conical state collapses above a critical field $H_{c2}$ and the field-polarized state is entered. 

In a small temperature interval just below $T_c$ the skyrmion lattice stabilizes  \cite{2009_Muhlbauer_Science}. In reciprocal space the skyrmion lattice represents approximately a triple-$\vec{Q}$ state for all field directions, where the sum over the wave-vectors vanishes forming a triangle. As the modulus $\vert \vec{Q} \vert$ is constant, the wave vectors enclose rigid angles of $120^{\circ}$. It is important to emphasize that the phase relationship of the three modulations is fixed and rigid as confirmed by small-angle neutron scattering~\cite{2011_Adams_PhysRevLett}. For the energetic stabilization of the skyrmion lattice phase fluctuation correction to the free energy are essential \cite{2009_Muhlbauer_Science, 2013_Buhrandt_PhysRevB}. Depending on field direction small changes of precise orientation of the skyrmion lattice with respect to the crystal lattice are qualitatively and quantitatively in excellent agreement with the cubic magnetocrystalline anisotropies up to sixth order in spin orbit coupling~\cite{2018_Adams_PhysRevLett}. 

It is interesting to note that cubic magneto-crystalline anisotropies may stabilize skyrmion lattice order for specific directions only. A second skyrmion phase of this type was recently identified in {\cso} at low temperatures and magnetic field along $\langle 100\rangle$~\cite{2018_Chacon_NatPhys}. The study reported here focusses instead on the high temperature skyrmion phase and the limit of very small cubic magneto-crystalline anisotropies relevant to MnSi.

The account of our key finding begins with a comparison of the quantitative value of the real and imaginary parts of the longitudinal and transverse AC susceptibility, as depicted in terms of the color shading of the magnetic phase diagram for field parallel $\langle100\rangle$ in Figs.~\ref{figure1}(a)--\ref{figure1}(d), before turning to typical data, as shown in Figs.~\ref{figure1}(e)--\ref{figure1}(i). The phase boundaries are thereby inferred from the longitudinal AC susceptibility ($\chi_{\mathrm{ac}}^{\mathrm{L}}$, squares), the susceptibility calculated from the magnetization ($\mathrm{d}M/\mathrm{d}H$, circles), and the specific heat (diamonds). Data inferred from temperature and field sweeps are indicated by light and dark colors, respectively. For what follows below it is important to point out the lobed phase boundary between the skyrmion lattice phase and the FD regime. A comprehensive account of the definitions of the phase transitions for the different quantities as well as the phase diagram has been reported in Refs.~\cite{2012_Bauer_PhysRevB, 2013_Bauer_PhysRevLett, 2016_Bauer_Book}. In comparison to previous work the information presented in the following on the longitudinal susceptibility is of much greater detail; rudimentary data of the transverse susceptibility may be found in Ref.\,\cite{2015_Chacon_PhysRevLett}.

The well-known real part of the longitudinal AC susceptibility, shown in terms of the color shading in Fig.~\ref{figure1}(a), is essentially constant in the helical phase (blue shading) and assumes an increased constant value in the conical phase (red shading). In the skyrmion lattice phase, the susceptibility is also constant at a lower value as compared to the conical phase~\cite{2017_Bauer_PhysRevB}. The color shading highlights particularly well the FD regime (green shading). 

In comparison, the real part of the transverse susceptibility, shown in Fig.~\ref{figure1}(b), displays an increase as a function of increasing magnetic field in the helical and conical phase (changing from blue to green shading), superimposed by a maximum at the helical-to-conical transition~\cite{2015_Chacon_PhysRevLett, 2015_Nii_NatCommun}. The details of this behavior are in excellent agreement with theoretical models of the magnetic order presently available and will be reported elsewhere~\cite{2017_Rucker_PhD}. In the skyrmion lattice phase the susceptibility is constant and enhanced. Of interest for the study reported here is the transverse susceptibility in the FD regime, which displays an enhanced value that is quantitatively comparable to the longitudinal susceptibility (yellow-green shading). The color shading illustrates that the longitudinal and transverse susceptibility are essentially isotropic in the FD regime and thus independent of the magneto-crystalline anisotropy.

Additional information on the nature of the magnetic response is provided by the imaginary part of the AC susceptibility shown in Figs.~\ref{figure1}(c) and \ref{figure1}(d), recorded simultaneously as the out-of-phase signal component. The imaginary part of the longitudinal AC susceptibility, depicted in Fig.~\ref{figure1}(c), displays a finite value characteristic of dissipation that is small at the helical-to-conical transition and prominently large at the conical-to-skyrmion lattice transition. This is consistent with the first-order nature of this transition as reported in detail in Refs.~\cite{2012_Bauer_PhysRevB, 2013_Bauer_PhysRevLett, 2014_Levatic_PhysRevB, 2016_Bauer_PhysRevB}.

The imaginary part of the transverse AC susceptibility, shown in Fig.~\ref{figure1}(d), assumes a finite value in the helical state when approaching the conical phase and in all of the conical phase. This reflects the dissipation due to changes of the propagation direction of the modulation vector under oscillating transverse fields. An additional enhancement is observed at the helical-to-conical transition that may be attributed to the transition itself. Interestingly, for the excitation frequency and amplitude used in our study, the imaginary part of the AC susceptibility is vanishingly small in the skyrmion lattice phase as well as in all of the paramagnetic state including the FD regime. This observation is consistent with the observation that the skyrmion lattice does not follow changes of field orientation in the limit of very small transverse excitation amplitude and frequency, as observed in kinetic SANS~\cite{2016_Muhlbauer_NewJPhys}. 

The magnetic field dependence of the AC susceptibility, shown in Figs.~\ref{figure1}(e)--\ref{figure1}(i), provides further information on the character of the FD regime at finite magnetic field. Shown in Fig.~\ref{figure1}(e) is the real part of the longitudinal susceptibility at selected temperatures as denoted in the figure (curves are shifted vertically for better visibility). The interest concerns here contributions reminiscent of the skyrmion lattice phase as illustrated in terms of the dark and light gray shading. It is important to emphasize that dark gray shading is used in the field range of the long-range ordered state (below $T_{c}$) and light gray shading in the FD regime (above $T_{c}$). For $T = 26.8$~K the reduced value of the susceptibility at zero field corresponds to the helical state, followed by the enhanced constant susceptibility in the conical state between ${\sim}0.1$~T and ${\sim}0.45$~T. The reduction of the susceptibility in the skyrmion lattice phase with respect to the surrounding conical phase at temperatures of 27~K and above is marked by dark gray shading. 

Of particular interest is the magnetic field dependence at $T = 28.8$~K, which begins in the helical state but enters the FD regime, while next cutting across the skyrmion lattice phase as a re-entrant state within the FD regime. This sequence of different states reflects the lobed phase boundary. The suppression of the susceptibility is here partly in the long-range ordered skyrmion lattice (dark gray shading) and partly in the FD regime (light gray shading). Yet, the susceptibility curve is basically featureless across the transitions between light and dark gray shading. Slightly increasing the temperature to 28.9~K, the field sweep still starts in the helical state, followed by the FD regime without a re-entrant segment in the skyrmion lattice phase. Here the FD regime features the reduced susceptibility at intermediate fields (light gray shading). As the helimagnetic order at zero magnetic field finally vanishes at $T_{c}$, just above 29~K, a reduced value of the susceptibility (light gray shading) survives as a signature within the FD regime. Closer inspection reveals the same suppression of the susceptibility in the color shading of the FD regime, shown in Fig.~\ref{figure1}(a). 

The evolution of the susceptibility for field parallel to $\langle100\rangle$ suggests the presence of skyrmionic correlations in the FD regime. As shown in Figs.~\ref{figure1}(f)--\ref{figure1}(h), the same qualitative and quantitative behavior is also observed for field along $\langle110\rangle$, $\langle211\rangle$, and $\langle111\rangle$, where the differences of the field range of the skyrmion lattice phase reflect the well-understood weak magnetocrystalline anisotropies that are fourth-order in spin--orbit coupling~\cite{1980_Bak_JPhysC, 2012_Bauer_PhysRevB, 2018_Adams_PhysRevLett}. This suggests, in turn, the presence of skyrmionic fluctuations in the FD regime under small applied magnetic fields (roughly between $\sim0.1\,{\rm T}$ and $\sim0.3\,{\rm T}$), regardless of the crystallographic direction in which the magnetic field is applied.

The possible existence of skyrmionic correlations in the FD regime is further corroborated by the transverse susceptibility, shown in Fig.~\ref{figure1}(i). Consistent with the longitudinal susceptibility as compared to the conical state, the real part of the transverse susceptibility features a considerable enhancement in the skyrmion lattice phase marked by dark gray shading. This enhancement evolves into an enhancement in the FD regime. Namely, between 28.6~K and 28.8~K the field sweep under increasing field exits from the skyrmion lattice phase into the FD regime, which displays also an enhanced susceptibility as compared to the susceptibility of the conical state (black line extrapolated from the behavior at 26.7~K). Interestingly, on top of this large enhancement there is a faint reduction, visible at 28.9~K, which vanishes at the highest temperatures. The shallow minimum around 0.2~T is reminiscent of the minimum in the longitudinal susceptibility in the FD regime, providing evidence for a response that is essentially isotropic.

%%%%%%%%%%%%%%%%%%%%%%%%%%%%%%%%%%%%%%%%%%%%%%%

\subsection{Small-Angle Neutron Scattering}
\label{SANS}

\begin{figure}
\includegraphics[width=1.0\linewidth]{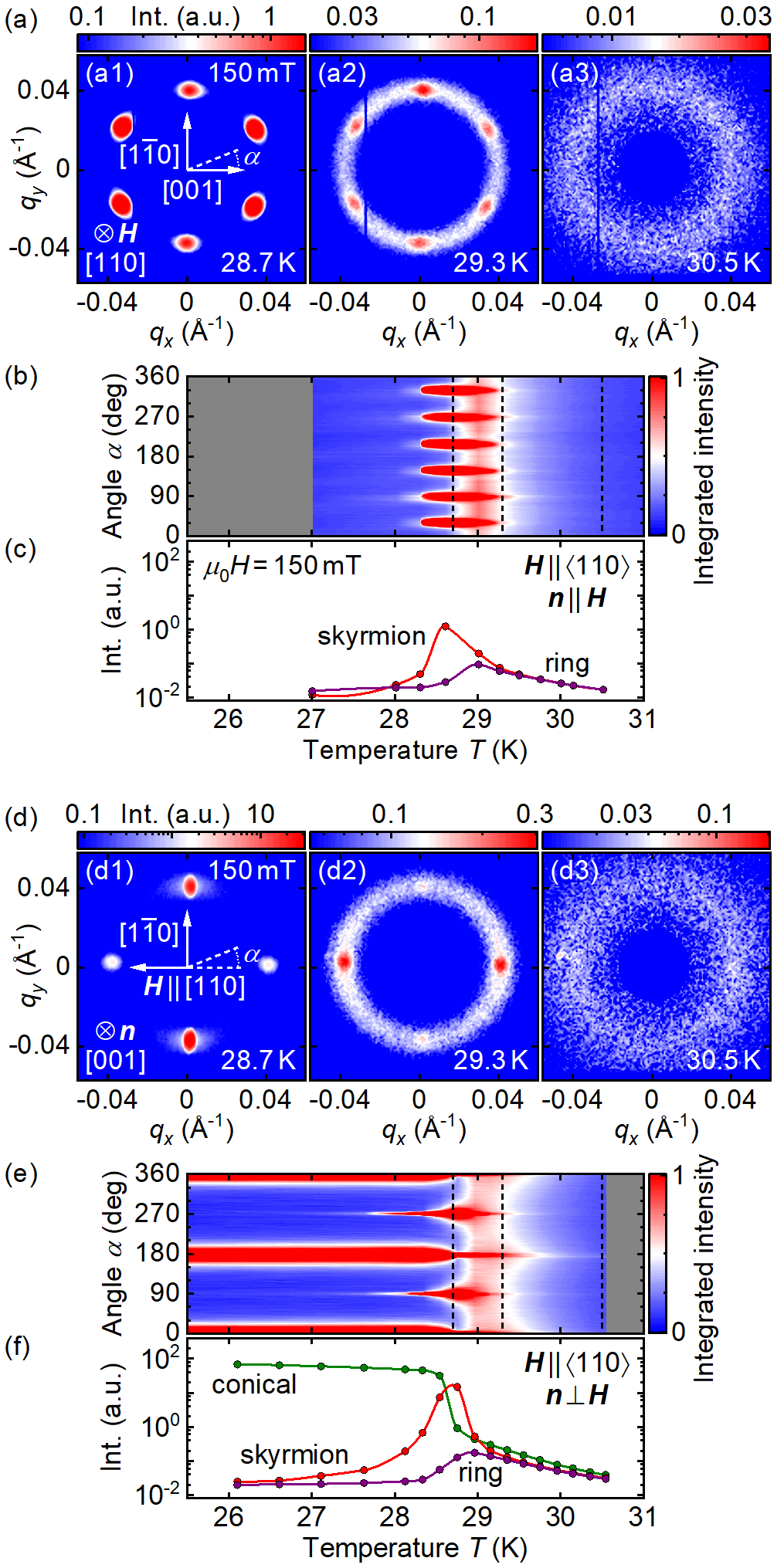}
\caption{Small-angle neutron scattering in an applied magnetic field parallel [\mbox{(a)--(c)}] and perpendicular [\mbox{(d)--(f)}] to the neutron beam. (a),(d)~Typical SANS patterns for temperatures within the skyrmion lattice state, just above $T_{c}$, and well above $T_{c}$ (from left to right). (b),(e)~Temperature evolution of the azimuthal SANS intensity distribution. Dashed vertical lines mark the temperatures shown in panels (a) and (d), respectively. (c),(f)~Temperature dependence of the elastic peak intensity characteristic of the conical state, the skyrmion lattice, and the fluctuation-disordered regime (ring) on logarithmic scale.}
\label{figure2}
\end{figure}

To obtain microscopic information on the nature of the magnetic correlations in the FD regime, small-angle neutron scattering (SANS) was performed. Shown in Fig.~\ref{figure2} is an overview of typical SANS data for magnetic field parallel [Fig.~\ref{figure2}(a)--\ref{figure2}(c)] and perpendicular [Fig.~\ref{figure2}(d)--\ref{figure2}(f)] to the neutron beam. A comparison of data recorded under field cooling and field heating showed the same behavior without evidence for thermal hysteresis in the scattering patterns. This is consistent with the bulk properties, which did not show any evidence for hysteresis in the skyrmion lattice phase and FD regime either (differences of domain populations in the helical state addressed in Ref.~\cite{2017_Bauer_PhysRevB} are not of interest here). Furthermore, the SANS measurements as well as the MIEZE spectroscopy reported in the next section were recorded for magnetic field along $\langle110\rangle$ as the corresponding plane normal to $\langle110\rangle$ contains all high-symmetry directions. The isotropic behavior of the FD regime observed in the susceptibility shown in Fig.~\ref{figure1}, where most data were recorded for $\langle100\rangle$, allows to connect the susceptibility with the SANS data.

Shown in Fig.~\ref{figure2}(a1) is the characteristic sixfold intensity distribution as recorded in the skyrmion lattice phase of MnSi at 28.7~K for an applied magnetic field of 0.15~T along $\langle110\rangle$ (cf. sum over a one-axis rocking scan as explained above). In the FD regime at 29.3~K and the same magnetic field, the SANS pattern comprises a ring of scattering intensity with a faint sixfold azimuthal intensity modulation reminiscent of the skyrmion lattice phase, as shown in Fig.~\ref{figure2}(a2). We note that a similar scattering pattern is shown in Fig.~1(a) of Ref.~\cite{2017_Pappas_PhysRevLett} (data at 29~K), where it was not discussed as this work had a different focus. This suggest the presence of skyrmionic correlations in the FD regime as a precursor phenomenon prior to the onset of long-range skyrmion lattice order. 
However, while the similarity with the skyrmion lattice phase is striking we note that the SANS pattern alone does not allow to distinguish between genuine triple-$\vec{Q}$ and fortuitous single-$\vec{Q}$ correlations under $120^{\circ}$. We return to this issue in section \ref{FMR}. When further increasing the temperature in the FD regime to 30.5~K, the putative skyrmionic correlations vanish, leaving behind the ring of scattering intensity as shown in Fig.~\ref{figure2}(a3).

The detailed evolution of the azimuthal intensity distribution as a function of temperature is shown in Fig.~\ref{figure2}(b), where the temperatures of the patterns shown in Fig.~\ref{figure2}(a) are marked as dashed lines and $\alpha$ is defined in Fig.~\ref{figure2}(a1) (data were binned in sections of $2^{\circ}$). This underscores that the sixfold azimuthal intensity variation in the FD regime has the same orientation as the skyrmion lattice phase. Comparison of various intensities shown in Fig.~\ref{figure2}(c) on a logarithmic intensity scale reveals that the sixfold pattern attributed to the skyrmion lattice is substantial at first in the FD regime and merges with the ring of intensity under increasing temperature. Thus, the sixfold scattering pattern represents an important facet of the FD regime when approaching the onset of long-range skyrmion lattice order.

The data recorded for field perpendicular to the neutron beam, shown in Figs.~\ref{figure2}(d)--\ref{figure2}(f), complements the behavior for field parallel to the neutron beam presented so far. For this configuration, intensity along the horizontal direction ($\alpha = 0$ and $\alpha = 180^{\circ}$) corresponds to conical order, whereas intensity along the vertical direction ($\alpha = 90^{\circ}$ and $\alpha = 270^{\circ}$) corresponds to the ring perpendicular to the field notably the six-fold maxima. Typical data in the skyrmion lattice phase at 28.7~K, shown in Fig.~\ref{figure2}(d1), comprise strong intensity for the skyrmion lattice and very weak intensity associated with conical correlations. In comparison, a ring of scattering intensity with additional maxima is observed in the FD regime at 29.3~K, shown in Fig.~\ref{figure2}(d2). In combination with the ring of scattering intensity seen for field parallel to the neutron beam, cf.\ Fig.~\ref{figure2}(a2), this implies scattering intensity on the entire surface of a sphere in the FD regime under the applied magnetic field of 0.15~T. 

In addition, maxima can be discerned in Fig.~\ref{figure2}(d2) at the top and bottom (faint) and as well as horizontally (strong) that are characteristic of skyrmionic and conical correlations, respectively. It is important to note the much larger phase space assumed by the intensity on the ring perpendicular to the field direction, as compared to their conical counterparts that are confined to regions on the sphere close to wavevectors parallel to the field. That is, when integrating over the entire ring perpendicular to the field a scattering intensity is found which is larger than the intensity attributed to the conical correlations.

At 30.5~K the data shown in Figs.~\ref{figure2}(a3) and \ref{figure2}(d3) provide evidence of intensity that is almost uniform on the surface of a sphere with a gentle enhancement in the field direction and the weak sixfold maxima seen in Figs.~\ref{figure2}(a2) and (d2) have vanished. Upon increasing field, this broad distribution of scattering intensity continuously changes, and at a field of 0.5~T it is practically concentrated close to two points along the field direction. The details of the magnetic field dependence of this redistribution and the question for the Brazovskii scenario under applied magnetic field are beyond the scope of the work reported here.

The temperature dependence of the azimuthal intensity variation of the data shown in Fig.~\ref{figure2}(d) is depicted in Fig.~\ref{figure2}(e). The pronounced scattering due to the conical state at low temperatures ($\alpha = 0$ and $\alpha = 180^{\circ}$) shifts into pronounced scattering characteristic of the skyrmion lattice phase. When further increasing the temperature and entering the FD regime, the intensity distribution on the surface emerges together with residual intensity characteristic of conical and skyrmionic correlations. This evolution is corroborated in Fig.~\ref{figure2}(f), which displays the peak intensities observed for the conical and skyrmionic correlations as well as the ring of scattering. At the highest temperature studied in the FD regime of 30.5~K the signatures for conical and skyrmionic correlations merge with the scattering on the surface of the sphere denoted as ring.

%%%%%%%%%%%%%%%

\begin{figure}
\includegraphics[width=1.0\linewidth]{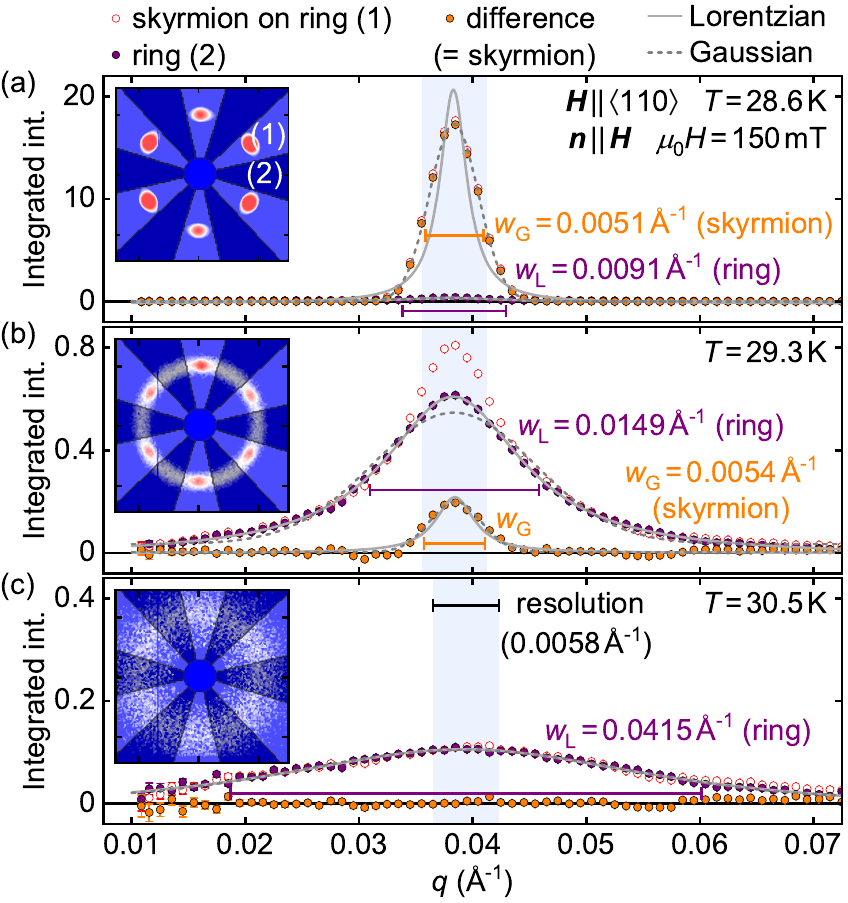}
\caption{Evolution of various radial correlation lengths as inferred from the SANS data at  $\mu_0 H=150\,{\rm mT}$. The instrumental resolution is marked by light-blue shading (a) Intensity as a function of $q$ in the skyrmion lattice phase at $T=28.6\,{\rm K}$, where data in the sectors marked (1) and (2) of the intensity pattern shown in the inset were summed up.  (b) and (c) Intensity as a function of $q$ in the FD regime at $T=29.3\,{\rm K}$ and the paramagnetic state outside the FD regime at $T=30.5\,{\rm K}$. Data in the sectors marked (1) and (2) are denoted skyrmion on ring and ring, respectively. The difference between sectors (1) and (2) is marked as skyrmion. See text for further details.}
\label{figure3}
\end{figure}

Shown in Fig.~\ref{figure3} is the intensity as a function of the modulus of $\vec q$ for the skyrmion lattice phase and the adjacent paramagnetic state when slightly increasing the temperature as inferred from the scattering patterns shown in Fig.~\ref{figure2}(a). The intensity represents sums of the sectors marked (1) and (2) as depicted in the insets, corresponding to the six-fold pattern and the ring between the spots, respectively. In the following these two directions are denoted with the subscripts $s$ and $r$, respectively. For what follows it is important to note that we expect a Lorentzian lineshape for scattering intensity that is not constrained by the experimental set up. In contrast, resolution limited scattering gives essentially rise to a Gaussian lineshape. The calculated experimental resolution of our set up taking into account the scattering geometry, wave length spread, and resolution of the detector is given by $2 \sigma_q = 0.0058\,{\rm \AA}^{-1}$. This corresponds to twice the standard deviation of the Gaussian and is depicted by the light-blue shading in Fig.~\ref{figure3}. Its full-width-half-maximum is given by $\Delta_q\, = 2 \sigma_q \sqrt{2 \log 2} \approx 0.0068\,{\rm \AA}^{-1}$. 

For the skyrmion lattice phase shown in Fig.~\ref{figure3}(a) we find a Gaussian lineshape in sector (1), that is characteristic of resolution-limited scattering intensity. To emphasize this point a Lorentzian lineshape is shown in addition. Quantitatively the measured Gaussian lineshape in sector (1) displays a full-width-half-maximum $w_{\rm G} = 0.0051\,{\rm \AA}^{-1}$ that is smaller than the calculated resolution $\Delta_q\,=0.0068\,{\rm \AA}^{-1}$ (see section \ref{methods} for details). 

The measured resolution limit $w_{\rm G}$ permits to provide a lower estimate of the correlation length of the faint sixfold scattering pattern. Namely, assuming that we can resolve a signal contribution of 20\,\% beyond the resolution limit the correlation length of such a signal would be $\xi_{\rm s} \gtrsim  2/\sqrt{(1.2 w_{\rm G})^2-w_{\rm G}^2} = 591\,{\rm \AA}$. In other words, the correlation length $\xi_{\rm s}$ associated with the six-fold intensity distribution must be considerably larger than $591\,{\rm \AA}$. This is consistent with the very large correlation lengths in the skyrmion lattice phase inferred from the magnetic mosaicity \cite{2011_Adams_PhysRevLett}.

While there is essentially no intensity in sector (2)  between the skyrmion peaks shown in Fig.~\ref{figure3}(a), it is nonetheless sufficient for a cautious assessment. Here we find a Lorentzian lineshape with a full-width-half-maximum $w_{\rm L} = 0.0091\,{\rm \AA}^{-1}$.  This corresponds to a correlation length $\xi_{\rm r} = 2/\sqrt{w_{\rm L}^2 - \Delta_q^2}\approx 330\,{\rm \AA}$. This value is consistent with the correlation length observed in zero magnetic field at the paramagnetic to helical transition as reported in Ref.\,\cite{2013_Janoschek_PhysRevB}. Importantly, it is significantly smaller than $\xi_{\rm s}$.

In the FD regime at 29.3\,K shown in Fig.~\ref{figure3}(b) the intensity in sectors (1) and (2) as depicted by open and filled symbols, respectively, displays the same $q$-dependence except within a small range in the vicinity of the maximum. Closer inspection reveals that the lineshape of the intensity in sector (2) is well described by a Lorentzian (the Gaussian is depicted by a dashed line). The full-width-half-maximum $w_{\rm L} = 0.0149\,{\rm \AA}^{-1}$ of the Lorentzian lineshape exceeds the resolution limit substantially amounting to a correlation length $\xi_{\rm r} \approx 150\,{\rm \AA}$.

The difference between the intensities in sectors (1) and (2), depicted by filled orange symbols, represents the faint six-fold intensity distribution in the FD regime. Qualitatively, the lineshape of this difference corresponds to a Gaussian that is characteristic of resolution-limited scattering. The associated full-width-half-maximum is $w_{\rm G} = 0.0054\,{\rm \AA}^{-1}$. Thus the faint sixfold intensity contribution in the FD regime features a correlation length substantially exceeding $\xi_{\rm s}\gtrsim 591\,{\rm \AA}$ as estimated in the skyrmion lattice phase, and may be as large as many thousand {\AA} \cite{1990_Pedersen_JApplCrystallogr}. 

At the border of the FD regime at 30.5\,K, shown in Fig.~\ref{figure3}(c), the sixfold intensity variation has vanished and the intensity in sectors (1) and (2) displays the same $q$ dependence. Here the lineshape is well-decribed by a Lorentzian with a full-width-half-maximum, $w_{\rm L} = 0.0415\,{\rm \AA}^{-1}$, that exceeds the instrumental resolution by a large margin. The associated correlation length is $\xi_{\rm r}\approx 49\,{\rm \AA}$.

The simple analysis of the intensity pattern of the paramagnetic state at 29.3\,K shown in Fig.~\ref{figure2}(d2) qualitatively and quantitatively establishes significant differences of typical correlation lengths of the ring and of the faint six-fold pattern in the FD regime. The ring is characterized by correlation lengths that are consistent with the behaviour at zero field as reported in Ref.\,\cite{2013_Janoschek_PhysRevB}. In contrast, both the skyrmion lattice order as well as the skyrmionic correlations in the FD regime display correlation lengths that are much larger and exceed the resolution limit. This represents a rather remarkable separation of scales in the plane perpendicular to the applied field, where the six-fold skyrmionic correlations in the paramagnetic FD regime extend over very large distances forming extended textures. The intensity that is evenly distributed on the ring perpendicular to the applied field could be either attributed to helical fluctuations or to skyrmionic fluctuations with shorter correlation lengths so that its six-fold pattern has not yet locked into the magnetocrystalline potential. Our SANS data does not allow to distinguish between these two possibilities. We return to this question at the end of section \ref{FMR}, where we present a strong plausibility argument for dominant skyrmionic correlations. A discussion of the consistency of the sixfold pattern in the FD regime with the underlying magnetocrystalline anisotropies may be found in section \ref{scales}.

%%%%%%%%%%%%%%%%%%%%%%%%%%%%%%%%%%%%%%%%%%%%%%%

\subsection{Neutron Resonance Spin Echo Spectroscopy}
\label{MIEZE}

The evidence for skyrmionic correlations in the FD regime as observed in the AC susceptibility and the resolution limited skyrmionic correlations observed in small-angle neutron scattering raises the question for their lifetime. This information proves to be essential for demonstrating that the SANS patterns are indeed skyrmionic as inferred from the FMR data reported below, as well as for unraveling the nature of the condensation of skyrmion lattice phase when starting from the paramagnetic regime. Although there is no evidence for hysteresis in any of the properties studied in the temperature and field range of interest here, we find it also helpful to confirm spectroscopically that there are no static remnants of the skyrmion lattice causing the sixfold intensity modulation in the FD regime. 

An ideal probe to clarify this question is neutron spin-echo spectroscopy, which offers the required ultra-high energy resolution. Considering the SANS intensity patterns shown in Fig.~\ref{figure2} we focused in our study on the location of the skyrmion lattice peaks. For a consistency check, we also performed test measurements at the location of the conical peaks for an applied magnetic field of 240\,mT (not shown). These measurements performed at RESEDA using the LMIEZE set-up were found to be in excellent agreement with the measurements performed at IN15 at the ILL reported in Ref.\,\cite{2017_Pappas_PhysRevLett}, where so-called ferromagnetic spin-echo spectroscopy was used.

\begin{figure}
\includegraphics[width=1.0\linewidth]{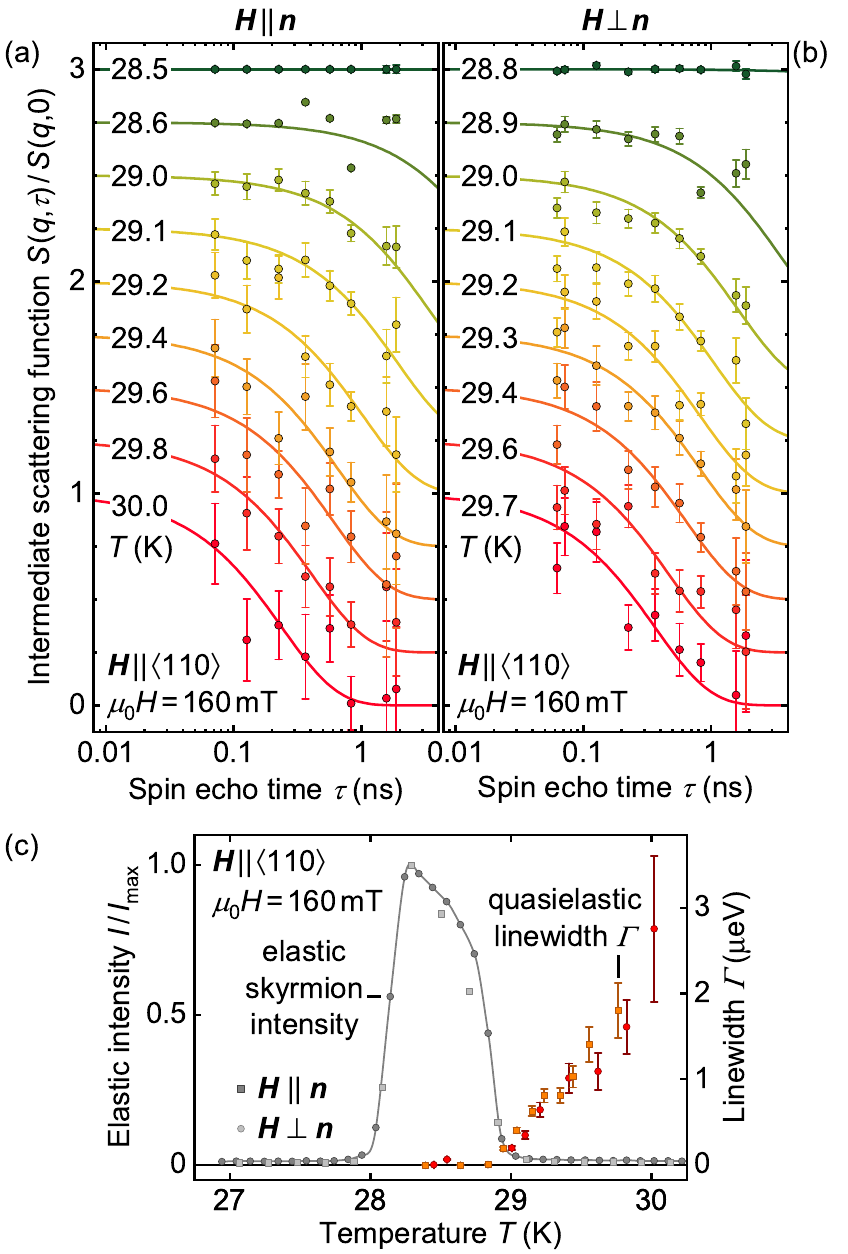}
\caption{Typical quasi-elastic behavior in the skyrmion lattice phase as determined by the MIEZE implementation of neutron resonance spin echo spectroscopy. (a),(b)~Normalized intermediate scattering function $S(q,\tau)/S(q,0)$ as a function of spin echo time $\tau$ for various temperatures below and above $T_{c}$ for magnetic field parallel and perpendicular to the neutron beam.  (c)~Elastic scattering intensity of the skyrmion lattice and quasi-elastic linewidth, $\Gamma$, both as a function of temperature. Error bars of the elastic intensity are smaller than the symbol size. The energy resolution of the quasi-elastic data are well below 0.1~$\mu$eV.}
\label{figure4}
\end{figure}

Shown in Fig.~\ref{figure4} are the results of the MIEZE neutron spin echo spectroscopy, where the directions of the applied magnetic field and the neutron beam were chosen such that the dynamical properties in the plane perpendicular to the applied magnetic field at one of the spots of the sixfold skyrmionic pattern was tracked. As for the SANS measurements, the magnetic field was applied along $\langle110\rangle$. Data were recorded under applied magnetic fields parallel and perpendicular to the neutron beam as denoted by squares and circles, respectively. Probing the same location in reciprocal space the same dynamical properties are expected, i.e., this test confirmed that the data were not contaminated by spurious scattering. 

Figures~\ref{figure4}(a) and \ref{figure4}(b) display the normalized intermediate scattering function $S(q,\tau)/S(q,0)$ as a function of the spin-echo time $\tau$ at various temperatures below and above $T_{c}$. For better visibility data are shifted vertically. Data for both constellations of field and beam direction are well described by a single exponential decay, $\exp(-(\Gamma /\hbar) \cdot \tau )$, where $\Gamma$ represents the quasi-elastic linewidth, and the associated lifetime is $t_0=\hbar/\Gamma$.

Summarized in Fig.~\ref{figure4}(c) is the temperature dependence of the elastic scattering intensity characteristic of the long-range ordered skyrmion lattice below $T_c \approx 28.8 $ K and the quasi-elastic linewidth, $\Gamma$, of the skyrmionic fluctuations observed above $T_c$. Both the elastic intensity and the MIEZE data were recorded at RESEDA in order to avoid any systematic differences of the temperature dependences of both quantities. In the skyrmion lattice phase the magnetic order is static within the tiny quasi-elastic energy resolution limit ($\ll 0.1~\mu$eV). The elastic intensity displays a very well-defined lower and upper transition temperature of the skyrmion lattice phase. 

Within experimental accuracy the quasi-elastic linewidth $\Gamma$ assumes a finite value above the temperature at which the elastic scattering intensity of the skyrmion lattice phase vanishes. The same quantitative temperature dependence of $\Gamma$ is observed for field parallel and perpendicular to the neutron beam as expected. Expressed in terms of the lifetime $t_0$ typical values corresponding to the SANS pattern at $29.3\,{\rm K}$ shown in Fig.~\ref{figure2}(a2) are roughly $t_0\approx 1\,{\rm ns}$ with $\Gamma\approx0.6\,{\rm \mu eV}$. When approaching the transition of the skyrmion lattice phase the lifetime increases and reaches at least several ns.  The remarkably small error bars and thus energy resolution clearly establish the dynamic character of the magnetic correlations in the FD regime. 

%%%%%%%%%%%%%%%%%%%%%%%%%%%%%%%%%%%%%%%%%%%%%%%

\subsection{Microwave Spectroscopy}
\label{FMR}

The similarities of the SANS scattering pattern of the long-range ordered skyrmion lattice below $T_c$ and the faint sixfold intensity pattern due to fluctuations in the FD regime above $T_c$ suggest that the latter possess already a skyrmionic character. However, our SANS data does not permit to distinguish between a multi-$\vec{Q}$ character of the fluctuations and fluctuating single-$\vec{Q}$ correlations that superpose fortuitously. The key aspect underlying the nontrivial topology of the skyrmions is a stiff phase relationship between Fourier components of the multi-$\vec{Q}$ state. This phase relationship has been demonstrated in SANS studies on bulk samples, where higher-order scattering could be tracked using so-called Renninger scans~\cite{2011_Adams_PhysRevLett}. The same phase relationship is also at the heart of the characteristic magnonic excitations observed in ferromagnetic resonance spectroscopy as well as inelastic neutron scattering.  

The excellent understanding of the interactions, magnetic order and magnetic phase diagram of cubic chiral magnets is reflected in a universal account of the collective spin excitations~\cite{1977_Date_JPhysSocJpn, 2012_Mochizuki_PhysRevLett, 2012_Onose_PhysRevLett, 2013_Okamura_NatCommun, 2015_Okamura_PhysRevLett, 2015_Schwarze_NatMater, 2017_Stasinopoulos_ApplPhysLett, 2017_Weiler_PhysRevLett, 2017_Garst_JPhysD}. Based on a few physically transparent parameters, the full spectrum of excitations, their spectral weight as well as the hybridization between modes were found to be in excellent agreement with experiment. In the helical and conical state two fundamental modes may be distinguished denoted $\pm Q$. In contrast, three fundamental modes exist in the skyrmion lattice phase, notably a clockwise (CW) and a counter-clockwise (CCW) mode, as well as a breathing mode (BM). All three modes are intimately related to the phase stiffness between Fourier components of the triple-$\vec{Q}$ state.

In the presence of long range magnetic order microwave spectroscopy represents a very well established and well understood technique. In comparison, microwave spectroscopy on dynamic correlations has been considered selectively only. The neutron spin echo spectroscopy clearly establishes a dynamic character of the skyrmionic correlations in the FD regime with characteristic lifetimes of the order of several $10^{-9}$\,s. It seems therefore safe to assume, that collective excitations may be detected by microwave spectroscopy whose periods are shorter than these lifetimes. 

\begin{figure}
\includegraphics[width=1.0\linewidth]{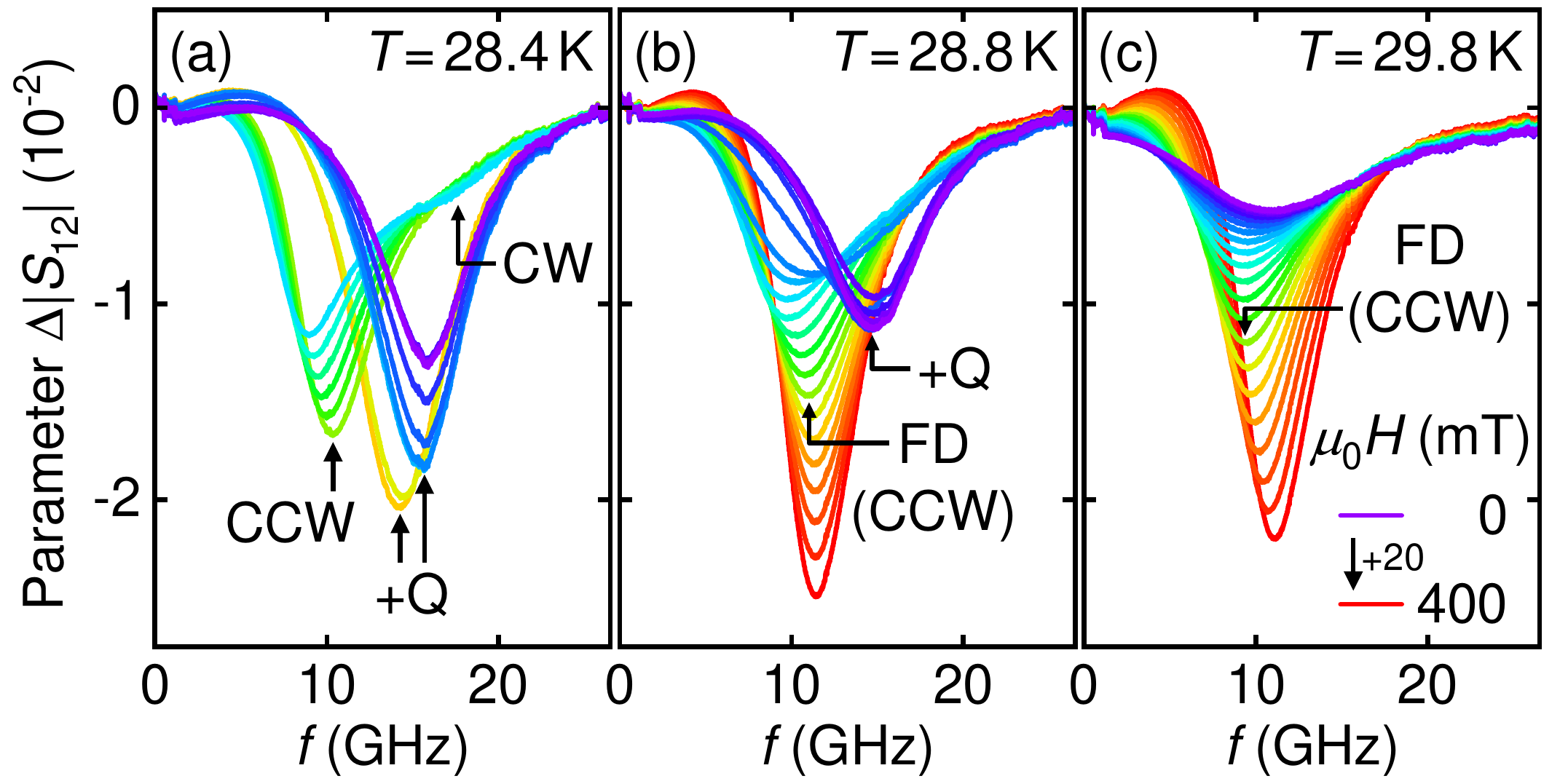}
\caption{Typical excitation spectra observed in all-electrical microwave spectroscopy, where the spectra correspond to values of the applied magnetic field between $\mu_0 H=0$ to 400\,mT in steps of 20\,mT. The color coding is identical in all panels and reflects the field value. In the conical state the $+Q$ mode is dominant; in the skyrmion lattice the clockwise (CW) and counter-clockwise (CCW) modes may be distinguished. (a)~Microwave spectra at 28.4\,K for selected magnetic fields from below to above the skyrmion lattice phase. (b) Microwave spectra at 28.8\,K. With increasing magnetic field data correspond to the helical state, the skyrmion lattice phase and the FD regime. (c)~Microwave spectra at 29.8\,K where all data are recorded within the FD regime. Spectra in the FD regime feature a pronounced CCW character.
}
\label{figure6}
\end{figure}

These considerations are clearly satisfied for typical microwave resonances in the range of $\sim10\,{\rm GHz}$ observed in our study. Typical data are shown in Fig.~\ref{figure6} for three selected temperatures and magnetic fields from $\mu_0 H=0$ up to 400\,mT in steps of 20\,mT. While the spectra appear to be broad, the resolution corresponds very well with seminal studies reported in the literature that allowed to discern different modes unambiguously, e.g., Ref.\,\cite{2013_Okamura_NatCommun} and Fig.\,5 in the supplement of Ref.\,\cite{2015_Schwarze_NatMater}. In particular, the line-shape is essentially Lorentzian and the resolution permits to identify and track the evolution of the excitations as a function of field under increasing temperatures above the ordering temperature.

Shown in Fig.~\ref{figure6}(a) are typical data in the ordered state at 28.4\,K, illustrating that different modes may be readily distinguished. For the sake of clarity, data for selected magnetic fields in a field-range encompassing the skyrmion lattice phase are displayed only. In the conical state below and above the skyrmion lattice phase the $+Q$ mode is dominant. In the skyrmion lattice phase the counter-clockwise mode is dominant, whereas the clock-wise mode possesses much less spectral weight resulting in a shoulder in $\Delta\vert S_{12}\vert$. Shown in Fig.~\ref{figure6}(b) are data recorded at 28.8\,K, where a magnetic field scan cuts across the helical state, the skyrmion lattice phase, and the FD regime. The shift of the resonance frequency from the $+Q$ mode at 16\,GHz to an excitation strongly reminiscent of the counter-clockwise mode at 10\,GHz is resolved well. Fig.~\ref{figure6}(c) was recorded at 29.8\,K and focusses on the FD regime. A well-developed single line is observed at the frequency of the counter-clockwise mode. 
 
\begin{figure}
\includegraphics[width=1.0\linewidth]{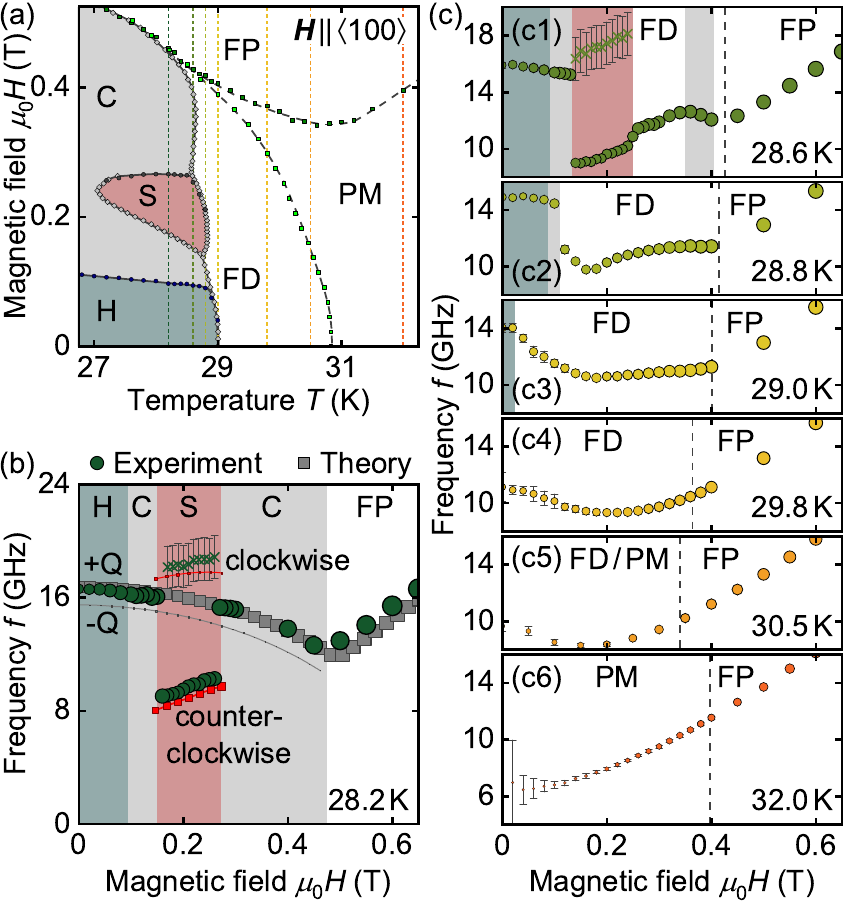}
\caption{All-electrical microwave spectroscopy. (a)~Phase dia\-gram for magnetic field along $\langle100\rangle$. Dashed vertical lines indicate temperatures for which data are shown. (b)~Typical experimentally observed (circles) and calculated (squares) excitation spectrum as a function of field for temperatures crossing the skyrmion lattice state well below $T_{c}$. The symbol size is a measure for the spectral weight of the mode. The background color shading indicates the magnetic state. (c)~Frequency and spectral weight of the dominant mode observed as a function of magnetic field for temperatures just below and above $T_{c}$.}
\label{figure7}
\end{figure}

The microwave spectra provide striking evidence for the presence of skyrmionic fluctuations in the FD regime under small applied magnetic fields. Shown in Fig.~\ref{figure7}(a) is the magnetic phase diagram as a function of applied magnetic field along $\langle100\rangle$ for the sample shape used in our microwave spectroscopy. It is again helpful to keep in mind the lobed phase boundary between skyrmion lattice phase and FD regime. Marked by vertical lines are the temperatures at which data were recorded as a function of magnetic field. The peak positions of these spectra are shown in Figs.~\ref{figure7}(b) and \ref{figure7}(c) for selected temperatures, where the color shading of the data points corresponds to the temperatures marked by the vertical lines in Fig.~\ref{figure7}(a), and the color shading in the background denotes the thermodynamic state. The size of the symbols corresponds to the weight of the modes as determined by integrating $\Delta|S_{12}|$ between 0.1~GHz and 27~GHz following subtraction of a reference signal. Thus, the value represents an experimentally extracted estimate of the spectral weight of the dominant mode for each field value. The error bars represent a conservative estimate of the accuracy at which the frequency of the minimum in $\Delta|S_{12}|$ may be determined taking into account the statistical noise of the signal.

The excellent account of the experimentally observed microwave spectra in terms of the excitation frequencies and spectral weight for the long-range ordered phases in the magnetic phase diagram of MnSi is illustrated in Fig.~\ref{figure7}(b) for 28.2~K~\cite{2015_Schwarze_NatMater}. The experimental data are depicted by circles, whereas the theoretical predictions are depicted by squares. As a function of magnetic field, the spectra are characteristic of the different magnetic phases, notably helical~(H), conical~(C), skyrmion lattice~(S), and field-polarized~(FP), where the phase boundaries are in excellent agreement with bulk properties notably magnetization, AC susceptibility, and specific heat. For the theoretical calculations the sample shape was taken into account, namely a thin disc, as the sample shape by virtue of the associated dipolar interactions leads to considerable changes of the weight and precise character of the excitation modes ~\cite{2015_Schwarze_NatMater}. 
	
Three prominent modes may be distinguished in the experimental data in terms of their qualitative field dependence and quantitative values as shown in Fig.~\ref{figure7}(b). First, in the helical and the conical state the $+Q$ mode dominates, decreasing monotonically from ${\sim}16$~GHz at zero field to ${\sim}12$~GHz at the onset of the field-polarized state. Second, in the skyrmion lattice phase the counter-clockwise gyration mode is dominant, whereas the clockwise mode is very faint (for the excitation geometry used, the breathing mode is vanishingly weak and not shown for clarity~\cite{2012_Onose_PhysRevLett, 2017_Stasinopoulos_SciRep}). As its key property, the counter-clockwise mode is much lower in frequency, increasing from ${\sim}8$~GHz to ${\sim}10$~GHz for increasing field. Third, in the field-polarized state the conventional Kittel mode is prominently observed, increasing linearly with increasing field.

Keeping in mind the key characteristics of the three main excitations, notably $+Q$, counter-clockwise, and Kittel mode, it is instructive to track the evolution of the magnetic field dependence when gradually increasing the temperature and entering the FD regime as summarized in six individual panels in Fig.~\ref{figure7}(c). Data points represent the minima of the spectra unless a second mode may be distinguished in terms of a shoulder as illustrated in Fig.\,\ref{figure6}\,(a). As in Fig.~\ref{figure7}(b) the field range of the long-range ordered phases (helical, conical, and skyrmion lattice) is depicted by the color shading in the background. The dashed vertical lines around ${\sim}0.4$~T represent the onset of the field-polarized state as inferred from Fig.~\ref{figure7}(a). 

At 28.6~K, shown in Fig.~\ref{figure7}(c1), the field sweep follows at first the behavior known from Fig.~\ref{figure7}(b). However, due to the lobed phase boundary between the skyrmion lattice phase and the FD regime, the field sweep exits from the skyrmion lattice phase into the FD regime between 0.25\,T and $0.35\,{\rm T}$ and re-enters the conical state between about 0.35\,T and 0.4\,T, before reaching the field-polarized state. In the conical phase the extracted resonance frequency (circles) exhibits a negative slope, $\mathrm{d}f/\mathrm{d}H<0$, characteristic of the $+Q$ mode. In the field range of the FD regime, the excitation frequency instead increases with increasing field, strongly reminiscent of the counter-clockwise mode of the skyrmion lattice phase. It is important to note that the changes of frequency between the different phases are abrupt.

The observation of a pronounced CCW collective character in the FD regime is corroborated by the field sweeps at 28.8~K and 29.0~K, shown in Figs.~\ref{figure7}(c2) and \ref{figure7}(c3). With increasing field the excitation frequency drops drastically from the helical/conical state, increases smoothly in the FD regime, and clearly changes its slope at the border to the field-polarized state. It is important to note that we track the minima in the spectra and that the lineshape remains essentially unchanged and well-defined. With increasing temperature data such as those shown in Fig.\,\ref{figure7}(c3) appear to suggest that the excitations in the conical state evolve continuously into those of the FD regime. However, as we detect a single pronounced minimum in this regime, it is hard to distinguish a single mode with mixed character from the possible presence of two independent modes with shifting weight that are close to each other.

Indeed, when further increasing the temperature of the sample above the zero-field helical-to-paramagnetic transition, shown in Figs.~\ref{figure7}(c4) and \ref{figure7}(c5), the field dependence of the excitation frequency evolves gradually into the characteristic behavior of a field-polarized paramagnet with a Kittel mode at large field as shown for 32~K in Fig.~\ref{figure7}(c6). Consistent with the paramagnetic state, the spectral weight at low fields is very low and the corresponding error bars are large. Such a cross-over notwithstanding, the data at 28.6\,K and 28.8\,K provide strong evidence of a dominant CCW skyrmionic character in the FD regime.

On this note, it is instructive to revisit the distribution of spectral weight observed in SANS in the FD regime, where the intensity on the surface of a sphere displays an enhancement in the plane perpendicular to the field and some enhancement in the direction of the field. While the sixfold intensity of the former may be attributed to skyrmionic correlations, it is not possible to distinguish skyrmionic and helical contributions elsewhere. Accordingly one might expect skyrmionic and helical/conical signal contributions in the FMR data. 

As stated in section \ref{methods}, in our set-up the static magnetic field was applied perpendicular to the CPW along the $z$ axis, whereas the strongest high-frequency component induced a precessional motion along the $x$-axis. Accordingly, our set-up was sensitive to signal contributions of skyrmionic correlations with $\vec Q$ vectors within the plane perpendicular to the static field, as well as conical correlations with $\vec Q$ vectors perpendicular to the $x$-axis (to excite the $\pm Q$ modes of the conical helix the ac field must be perpendicular to the modulation direction $\vec Q$).

With these aspects in mind, it is important to note that there is no evidence for collective modes characteristic of the conical state in the FD regime as even a small volume fraction of the $+Q$ mode in Fig.\,\ref{figure6}(b) would result in a pronounced shoulder if present in the FD regime. There are two possible explanations for this observation. First, the correlation length of the helical/conical fluctuations in the FD regime may be too short for their collective modes to develop even though their lifetimes are comparable to that of the skyrmionic fluctuations. Second, the correlations that contribute to the microwave spectra may be entirely skyrmionic in character. Given the quantitative size of the skyrmionic signal, the latter seems to be most likely. In turn, we conclude that the volume fraction of skyrmionic correlations in the FD regime is much larger than may be expected of the faint sixfold SANS intensity pattern.

Thus, for increasing temperature the magnetic field dependence of the microwave spectra is dominated by an excitation in the FD regime that is reminiscent of the counter-clockwise mode of the skyrmion lattice phase. As the lifetime of the skyrmionic correlations in the FD regime is much longer than the excitation frequencies, this demonstrates a triple-$\vec{Q}$ character of the fluctuations with a stiff phase relationship and, therefore, a pronounced skyrmionic content in the fluctuation spectrum.

%%%%%%%%%%%%%%%%%%%%%%%%%%%%%%%%%%%%%%%%%%%%%%%%%

\section{Discussion}
\label{discussion}

\subsection{Summary of key observations} 
\label{summary}

Our study concerns the nature of the paramagnetic state at the border of long-range skyrmion lattice order and the associated phase transition. Starting with the longitudinal and transverse susceptibilities we find signal contributions in the FD regime of the paramagnetic state that are remnants of long-range skyrmion lattice order. SANS in this regime reveals scattering intensity on the surface of a sphere, with increased weight in the plane perpendicular to the magnetic field  as well as along the field direction. The additional weight within the plane contains a six-fold intensity pattern, that is strongly reminiscent of skyrmion lattice order in three ways. First, the sixfold pattern features the same modulus $\vert\vec{Q}\vert$ as the long range skyrmion lattice order below $T_c$. Second, the orientation of the six-fold pattern with respect to the crystallographic lattice is unchanged the same as below $T_c$. 
Third, the correlation lengths and line shapes are resolution limited and clearly reminiscent of long range skyrmion lattice order.  

In order to confirm that these six-fold signatures in the paramagnetic FD regime are due to dynamic fluctuations and are clearly distinct from genuine static long-range order, we performed ultra-high resolution neutron resonance spin-echo spectroscopy. We find that the signal contributions at the location of skyrmion spots in the plane perpendicular to the applied field are dynamic down to $T_c$. They are characterized by a critical slowing down with lifetimes exceeding several 
10$^{-9}$s.

Moreover, to distinguish experimentally between single-$\vec{Q}$ and generic multi-$\vec{Q}$ correlations we used microwave spectroscopy. We find signatures in the FD regime of the counter-clockwise gyration mode reminiscent of the skyrmion lattice state. The frequency of this excitation of around 10\,GHz is much faster than and therefore consistent with the lifetime of the fluctuations. This provides strong evidence of multi-$\vec{Q}$ correlations with an underlying rigid phase relationship that is reminiscent of long-range skyrmion lattice order and the associated non-trivial topological winding.

It is helpful to note that it is difficult to distinguish different volume fractions of the conical and skyrmion lattice fluctuations in the SANS data, since the variation of the intensities is rather gradual and because only one-axis rocking scans could be performed. However, taking into account the large phase space of the surface of the sphere as well as the scattering in the plane perpendicular to the applied field, both contributions are clearly much stronger than the scattering in the field direction. Moreover, the microwave spectra in the FD regime are clearly dominated by the counter-clockwise modes consistent with a dominant volume fraction of the skyrmionic correlations in the FD regime under small applied magnetic fields.

\subsection{Consistency with the hierarchy of scales}
\label{scales}

The skyrmionic fluctuations in the FD regime are strongly reminiscent of the long range skyrmion lattice order in terms of the modulus of wavevectors, the orientation of the six-fold scattering pattern and the correlation lengths observed experimentally. From the analysis of the static long-range skyrmion lattice order \cite{2009_Muhlbauer_Science,2018_Adams_PhysRevLett} it is known that the orientation is determined by magneto-crystalline anisotropies that are sixth order in spin-orbit coupling and rather weak in MnSi. In this respect, it is at first sight surprising that the dynamic skyrmionic fluctuations observed above $T_c$ are oriented in a similar manner than the order below $T_c$. In order to enable the fluctuating skyrmionic patches to accommodate with the magneto-crystalline potential, large correlation lengths are required as estimated in the following.

Several experimental studies have established an internally consistent quantitative account of the magneto-crystalline anisotropies of MnSi. This includes in particular the paramagnetic to helimagnetic transition in zero magnetic field~\cite{2013_Janoschek_PhysRevB, 2008_Janoschek_PhD}, the helical to conical transition for different crystallographic directions~\cite{2017_Bauer_PhysRevB} and the precise orientation of long range skyrmion lattice order~\cite{2018_Adams_PhysRevLett, 2015_Adams_PhD}. 
Denoting the strength of spin-orbit coupling as $\lambda_{\rm SOC}$, the strength of the Dzyaloshinskii-Moriya interaction is linear in spin-orbit coupling, $D\sim\lambda_{\rm SOC}$. In comparison, the leading order cubic magneto-crystalline anisotropies are fourth order in spin-orbit coupling $J_{\rm cub,hel} \sim \lambda_{\rm SOC}^4$, determining the direction of the helical modulation. The  contributions to the magneto-crystalline anisotropies which control the in-plane orientation of the skyrmion lattice are sixth order in spin orbit coupling, $J_{\rm cub,SkL}\sim \lambda_{\rm SOC}^6$.

The hierarchy of energy scales may be expressed in terms of a hierarchy of length scales, following the strategy and notation introduced in the context of the paramagnetic to helical transition in zero magnetic magnetic field~\cite{2013_Janoschek_PhysRevB, 2008_Janoschek_PhD}. Considering the paramagnetic state at high temperatures and small fields the correlation length is short and dominated by ferromagnetic exchange interaction $J$. As the correlation length $\xi$ increases with decreasing temperature, the much weaker Dzyaloshinskii-Moriya interaction affect the character of the fluctuations when $\xi$ reaches values comparable to $\xi_{\rm DM}=1/Q\approx 26\,{\rm \AA}$ where $Q=D/J \approx0.039\,{\rm \AA}^{-1}$ is the magnitude of the modulation wavevector. As the correlation length $\xi$ increases further under decreasing temperature, the leading-order magneto-crystalline anisotropies begin to become important when $\xi \gtrsim \xi_{\rm cub, hel}$ where $\xi_{\rm cub, hel}^2 \propto 1/J_{\rm cub,hel} \propto 1/\lambda_{\rm SOC}^4$. When the correlation length exceeds $\xi > \xi_{\rm cub, hel}$, the helimagnetic fluctuations will start to favour $\langle111\rangle$ crystallographic directions in MnSi. From neutron scattering data in zero field \cite{2013_Janoschek_PhysRevB}, this length scale was determined to be $\xi_{\rm cub, hel}\approx 170\,{\rm \AA}$. This yields an estimate of the strength of spin-orbit coupling in MnSi $\lambda_{\rm SOC} \sim \xi_{\rm DM}/\xi_{\rm cub, hel} \sim 0.15$.

With the information above we are in the position to estimate the  length scale $\xi_{\rm cub, SkL}$, which the correlation length of the skyrmionic fluctuations should exceed in order to lock into the magneto-crystalline potential. Using $\xi_{\rm cub, SkL}^2 \propto 1/J_{\rm cub,SkL} \propto 1/\lambda_{\rm SOC}^6$ we obtain $\xi_{\rm cub, SkL} \sim \xi_{\rm cub,hel}/\lambda_{\rm SOC} \sim 1133\,{\rm \AA}$. In other words, as the correlation length reaches $\xi \gtrsim \xi_{\rm cub, SkL}$, the correlations will become sensitive to magneto-crystalline anisotropy terms that are sixth order in spin orbit coupling, $J_{\rm cub,SkL}\sim \lambda_{\rm SOC}^6$. 

We note that correlations with a skyrmionic triple-$\vec{Q}$ character, due to their inherent six-fold symmetry, are not sensitive to the leading order magnetocrystalline anisotropy $J_{\rm cub,hel}\sim \lambda_{\rm SOC}^4$ which possess a four-fold symmetry. Instead, the orientation of the skyrmionic triple-$\vec{Q}$ fluctuations are determined by the same mechanism that also fixes the in-plane orientation of the static long-range ordered skyrmion lattice below $T_c$.

The rough estimates of the correlation lengths inferred from the SANS data are perfectly consistent with these considerations. The faint sixfold intensity pattern displays a resolution limited radial correlation length that must exceed $\sim$591\,\AA\, substantially. Here it is interesting to note, that the correlation lengths of long-range skyrmion lattice order inferred from the mosaicity are exceptionally large, reaching resolution-limited values in excess of $\mu{\rm m}$, i.e., the long-range crystalline character of the skyrmion lattice is much better developed than of the helical state~\cite{2011_Adams_PhysRevLett}. Therefore, it seems plausible, that skyrmionic fluctuations in the paramagnetic state feature also very large correlation lengths exceeding the correlation length of conventional helimagnetic fluctuations by a large margin.

It is further interesting to note that the correlation length of 330\,\AA\, observed on the ring within the plane perpendicular to the field, see Fig.~\ref{figure3}(b), exceeds the length scale, $\xi_{\rm cub, hel}\approx 170\,{\rm \AA}$, associated with cubic magneto-crystalline that is fourth order in spin orbit coupling. Yet, we do not find enhanced scattering intensity in any of the $\langle 111\rangle$ crystallographic directions. This may be explained with the presence of the magnetic field, which exceeds typical values of the helical to conical transition at $H_{\rm c1}$. In turn we conclude that the intensity distribution on the surface of a sphere observed in our SANS studies comprises skyrmionic and conical fluctuations, where the former are sensitive to the sixth order spin-obit coupling terms and the orientation of the latter is governed by the magnetic field. As discussed above, the microwave spectra suggest, that the correlations in the FD regime under applied magnetic fields is dominated by skyrmionic fluctuations.

\subsection{Implications for the emergent electrodynamics}
\label{emergent}

The interplay of skyrmions with spin currents may be described by means of an emergent electrodynamics that accounts for non-vanishing Berry phases, where the presence of a skyrmion is described by a fictitious magnetic flux of one quantum per skyrmion \cite{2009_Neubauer_PhysRevLett, 2012_Schulz_NatPhys}. As a direct consequence of the non-trivial topology, skyrmions give rise to an additional contribution of the Hall signal, referred to as topological Hall effect. In the emergent electrodynamics the creation and destruction of skyrmions by virtue of a locally vanishing magnetization, also known as Bloch points, may be interpreted in terms of magnetic monopoles supporting one quantum of emergent magnetic flux~\cite{2013_Milde_Science}. 

Our studies shed new light on the results of high pressure studies of MnSi, where the helimagnetic transition is suppressed above a critical pressure of $p_c=14.6\,{\rm kbar}$. When approaching $p_c$ an anomalous $T^{3/2}$ temperature dependence of the resistivity and a topological Hall signal emerge in the paramagnetic metallic state \cite{2001_Pfleiderer_Nature,2003_Doiron_Nature,2007_Pfleiderer_JLTP,2013_Ritz_PhysRevB,2013_Ritz_Nature,2013_Chapman_PhysRevB}. Moreover, elastic neutron scattering revealed a broad distribution of scattering intensity on the surface of a sphere, referred to as partial magnetic order \cite{2004_Pfleiderer_Nature}. Additional muon-spin-rotation measurements and NMR \cite{2004_Yu_PhysRevLett,2007_Uemura_NatPhys}, which failed to detect a signal, suggested that this partial magnetic order represents spin textures that are dynamic on time scales between $10^{-10}\,{\rm s}$ and $10^{-11}\,{\rm s}$. A similar topological Hall signal has also been reported in {\mfs} \cite{2014_Franz_PhysRevLett}, suggesting similarities with an additional importance of defects and disorder.

Taken together the transport properties, elastic neutron scattering, mu-SR and NMR reported in the literature provide circumstantial evidence of the formation of some form of dynamical topological spin textures in MnSi at high pressures. However, the nature of these textures has been unresolved. Speculations included, for instance, a liquid of particle-like skyrmions or some form of three-dimensional textures~\cite{2006_Binz_PhysRevLett, 2006_Binz_PhysRevB}. Our study establishes that, at least in the presence of a small magnetic field, skyrmionic textures may form in the paramagnetic phase at ambient pressure with characteristic length scales exceeding several $10^3\,{\rm \AA}$ and lifetimes up to several $10^{-9}\,{\rm s}$. Thus, the spatial extend and the lifetimes are large as compared to typical charge carrier mean free paths as well as their scattering times. 

Up to now it was not possible to discern a topological Hall signal in the FD regime at ambient pressures. The topological Hall signal observed at finite pressures however shows a strong suppression of the signal with increasing temperature \cite{2013_Ritz_PhysRevB,2013_Ritz_Nature}. This reduction may be attributed to a combination of mechanisms, of which interband scattering may be most important. Taking additionally into account the rather narrow temperature range of the FD regime and the strong temperature dependence of the resistivity (which affects the anomalous Hall contributions), it seems technically not possible to identify a topological Hall contribution unambiguously at ambient pressures, although it might, in principle, be generated by the skyrmionic fluctuations.

\subsection{Nature of the skyrmion lattice transition}
\label{weak-cryst}

Turning to the skyrmion lattice transition, the evidence for particle-like skyrmions in cubic chiral magnets comprises the observation of individual skyrmions and  skyrmion clusters \cite{2010_Yu_Nature, 2017_Muller_PhysRevLett, 2018_Loudon_PhysRevB}, nematic skyrmion textures~\cite{2018_Huang_ArXiv}, as well as hard-spheres-like defects in dynamically rotating skyrmion domains by means of Lorentz transmission electron microscopy (LTEM)~\cite{2014_Mochizuki_NatMater, 2017_Pollath_PhysRevLett}. However, LTEM measurements require thin bulk samples for which additional energy scales generate substantial changes of the magnetic phase diagram~\cite{2011_Yu_NatMater}. In comparison, neutron scattering provides strong evidence of wave-like skyrmion lattice order in MnSi and related compounds. Higher harmonics as weak as $10^{-4}$ and resolution limited correlation lengths are characteristic of very smooth, harmonic long-range order. The wave-like character of skyrmion lattice order may also be inferred from the spectrum of collective excitations~\cite{2012_Onose_PhysRevLett, 2015_Schwarze_NatMater, 2017_Weiler_PhysRevLett}.

Returning to the possible scenarios of the paramagnetic to skyrmion lattice transition mentioned in the introduction, our study clearly rules out a transition without any precursor phenomena in the paramagnetic state. Given the pronounced effect of skyrmionic fluctuations with its six-fold scattering pattern in the paramagnetic FD regime and the weak first-order character of the transition from the paramagnet to long-range skyrmion lattice order, it appears that the formation of the skyrmion lattice can be interpreted in the framework of weak crystallization.

The Ginzburg-Landau functional for the magnetization field generically contains a quartic term, $\int d\vec r (\vec M^2)^2$ stabilizing magnetic order. In the presence of a finite magnetic field, a finite uniform component of the magnetization $\vec M_u$ will be induced. Replacing one of the four fields in the quartic term with this uniform component, one obtains a coupling of $\vec M_u$ to the modulated components of the magnetization whose form in Fourier space is given by  \cite{2006_Binz_PhysRevLett, 2006_Binz_PhysRevB, 2009_Muhlbauer_Science}
\begin{equation} \sum_{\vec{q}_1,\vec{q}_2,\vec{q}_3} (\vec{M}_u \cdot \vec{m}_{\vec{q}_1})
(\vec{m}_{\vec{q}_2}\cdot \vec{m}_{\vec{q}_3}) \delta_{\vec{q}_1+\vec{q}_2+\vec{q}_3,0},\label{cubic}
\end{equation}
where $\vec{m}_{\vec{q}}$ is the Fourier transform of $\vec{M}(\vec r)$ and the Kronecker delta ensures momentum conservation. This term is effectively cubic in the Fourier components $\vec m_{\vec{q}}$ with finite $\vec{q}$ representing the modulation of the magnetization. Moreover, as the competition between Dzyaloshinskii-Moriya interaction $D$ and the symmetric exchange interaction $J$ favours modulation vectors with a finite length $Q=D/J \approx0.039\,{\rm \AA}^{-1}$, this cubic term can lower the energy provided that different Fourier components are combined whose wavevectors add up to zero, i.e., they form triangles. Due to symmetry, the triangular wavevector configuration is confined to the plane perpendicular to the uniform component $\vec M_u$, i.e., to the applied magnetic field. This eventually favours the formation of a hexagonal skyrmion lattice perpendicular to the field.

The qualitative form of the contributions in the free energy that drive the formation of skyrmion lattice order are in strong analogy with the Landau soft-mode mechanism of the formation of crystals out of the liquid state. 
In the framework of the theory of weak crystallization, it is assumed that the crystallization transition is either second order or only weakly first order so that the free energy functional can be expanded in terms of the oscillatory components of the density, $\rho_{\vec{q}}$. Cubic interactions of this density play a special role, which in momentum space may be written as
\begin{equation}
\sum_{\vec{q}_1, \vec{q}_2, \vec{q}_3} \rho_{\vec{q}_1}
\rho_{\vec{q}_2} \rho_{\vec{q}_3}
\delta_{\vec{q}_1+\vec{q}_2+\vec{q}_3,0}.
\end{equation}
The ordered state gains energy from this term only when three ordering vectors of the crystal structure add up to zero, thus forming triangles, which in two spatial dimensions favours hexagonal lattices.

The striking analogy with the Landau soft-mode mechanism classifies the onset of skyrmion lattice order as a weak crystallization process. This is consistent with the Brazovskii scenario of a fluctuation-induced first order transition at zero field, which originates in an increase of phase space for the fluctuations in combination with mode-mode interactions. Increasing the magnetic field the fluctuations are quenched and a tricritical point (TCP) is reached around 0.4\,T~\cite{2013_Bauer_PhysRevLett, 2014_Nii_PhysRevLett, 2017_Pappas_PhysRevLett, 2019_Janoschek_PhysRevLett} where the transition to conical long-range order turns second-order. The application of a field provides a static uniform component of the magnetization, $\vec{M}_u$, that favours triangular wavevector configurations eventually stabilizing skyrmion lattice order at intermediate magnetic fields. Even though the associated triple-$\vec Q$ fluctuations are dynamic in the paramagnetic state, they already entail the topological signatures that are characteristic of the long-range ordered skyrmion lattice. 

%%%%%%%%%%%%%%%%%%%%%%%%%%%%%%%%%%%%%%%%%%%%%%%%%
\section{Conclusions}
\label{conclusions}

In summary, combining measurements of the ac susceptibility, small angle neutron scattering, neutron resonance spin echo spectroscopy and microwave spectroscopy we find evidence for fluctuating skyrmion textures with characteristic length scales exceeding $10^3\,{\rm \AA}$ and lifetimes up to several $10^{-9}\,{\rm s}$. The paramagnetic to skyrmion lattice transition shares remarkable analogies with the Landau soft-mode mechanism of the weak crystallization of liquids. Our observations in the paramagnetic state establish that key signatures of the non-trivial topological character of the skyrmion lattice order are already present in the paramagnetic state when approaching the transition temperature in small fields. 

\acknowledgements
We wish to thank T.\ Adams, S.\ Mayr, and P. B\"oni for fruitful discussions and assistance with the experiments. Financial support through DFG TRR80 (From electronic correlations to functionality, projects E1, F7; project number 107745057), DFG FOR960, DFG SPP2137 (PF 393/19, project number 403191981), ERC Advanced Grant 291079 (TOPFIT), ERC Advanced Grant 788031 (ExQuiSid), and BMBF Project 05K16WO6 is gratefully acknowledged. J.K, I.S., F.H., F.R., and A.C.\ acknowledge financial support through the TUM graduate school. F.X.H. was supported through a Hans Fischer fellowship of the Technical University of Munich -- Institute for Advanced Study of Marc Janoschek, funded by the German Excellence Initiative and the European Union Seventh Framework Programme under Grant agreement No. 291763. M.G. is supported by DFG SFB1143 (Correlated Magnetism: From Frustration To Topology, project A07), DFG Grant No. 1072/5-1 and DFG Grant No. 1072/6-1. 

%\bibliography{MnSiSkyrmionFluctuations}
%\bibliography{ExportedItems}


%%%%%%%%%%%%%%%%%%%%%%%%%%%%%%%%%%%%%%%%%%%%%%%%%

%\section{Supplement}

\end{document}